\documentclass[10pt,floatfix,aps,pra,amsfonts,twocolumn]{revtex4-1}
\usepackage{amsmath,amssymb,bm,amsthm}
\usepackage{graphicx}

\newcommand{\Eq}[1]{Eq.~(\ref{#1})}
\newcommand{\Sec}[1]{Section~\ref{Sec:#1}} 
\newcommand{\Fig}[1]{FIG.~\ref{Fig:#1}}
\newcommand{\ket}[1]{| #1 \rangle}
\newcommand{\bra}[1]{\langle #1 |}
\newcommand{\Av}[1]{\langle #1 \rangle} 

\newcommand{\Id}{\openone}
\newcommand{\mat}[1]{\begin{pmatrix}#1\end{pmatrix}}
\newcommand{\Lvt}{{\mathsf L}}  
\newcommand{\CFtr}[1]{\mathsf{\Upsilon}\!_{#1}} 
\newcommand{\oCFtr}[1]{{\mathsf \Upsilon}^\circ_{\!#1}}  

\newcommand{\CC}{\mathbb{C}}
\newcommand{\RR}{\mathbb{R}} 
\newcommand{\VV}{\mathbb{V}}
\newcommand{\fld}{\mathbb{F}}
\newcommand{\Quat}{\mathbb{H}}
\newcommand{\mi}{\imath} 

\newcommand{\Clf}{\bm{C\mspace{-2mu}l}}
\newcommand{\UCl}{\bm{C\!\ell}} 
\newcommand{\e}{\mathfrak{e}}
\newcommand{\ttg}{\tilde\e} 
\newcommand{\sep}{;} 
\newcommand{\op}{\hat}     
\newcommand{\xt}{\check}
\newcommand{\xtop}{\check}
\newcommand{\btg}{\xt\e} 
\newcommand{\stub}{\op{\mathfrak{r}}}
\newcommand{\gj}{{\tilde\jmath}} 

\newcommand{\am}{\mathfrak{a}}
\newcommand{\btq}{\op{\mathfrak{h}}} 
\newcommand{\ant}{\xtop{\am}}
\newcommand{\an}{\op\am}
\newcommand{\oa}{\op{a}}  
\newcommand{\anx}[2]{\oa_{#1\triangleleft #2}}
\newcommand{\sbtr}{{\bm\varsigma}} 
\newcommand{\pmi}{\iota}
\newcommand{\vac}{\varnothing}
\newcommand{\Sym}{\varSigma}
\newcommand{\Asym}{\varLambda}
\newcommand{\nm}{\mathfrak{n}}
\newcommand{\num}{\op{\nm}}
\newcommand{\Num}{\op{\mathcal{N}}}
\newcommand{\St}{\mathsf{\Xi}}
\newcommand{\xSt}{\mathsf{\xt\Xi}}
\newcommand{\xnum}{\xt{\mathfrak{n}}}
\newcommand{\xNum}{\,\xt{\!\mathcal{N}}}
\newcommand{\mdiv}{\mathop{\mathrm{div}}}

\newcommand{\un}[1]{\underline{#1}}
\newcommand{\mt}[1]{\mathsf{#1}}
\newcommand{\exor}{\mathbin{\bigcirc\mspace{-16mu}{+}\,}}

\newcommand{\xst}[1]{\xt{\un{#1}}}
\newcommand{\spn}{\mathcal}
\newcommand{\spU}{\spn U}
\newcommand{\spH}{\spn H}
\newcommand{\sps}{\op{\mathfrak{s}}}
\newcommand{\xsUU}{\xt\spU_{\mt U}}
\newcommand{\ve}{\mathbf}
\newcommand{\cla}{\mathfrak{a}}
\newcommand{\clb}{\mathfrak{b}}

\newcommand{\onz}{\op{n}^{\!\mathnormal 0}}
\newcommand{\xz}{\breve}
\newcommand{\Xr}[1]{\exor_{\!#1}}
\newcommand{\anxr}[2]{\oa_{#1\Xr{#2}}} 
\newcommand{\term}{\mathcal{T}}
\newcommand{\prds}[2]{\op{\mathfrak{s}}^{#1}_{#2}}
\newcommand{\prdz}[1]{\prds{z}{#1}}

\newcommand{\citeref}[1]{Ref.~\cite{#1}}
\newcommand{\sgm}{\op{\sigma}}

\begin{document}

\title{Clifford algebras, Spin groups and qubit trees}

\author{Alexander Yu.\ Vlasov}

\sloppy

\begin{abstract}
Representations of Spin groups and Clifford algebras derived from the structure of qubit trees are introduced in this work. 
For ternary trees the construction is more general and reduction to binary trees is formally defined by deletion of superfluous
branches. The usual Jordan--Wigner construction also may be formally obtained in this approach by bringing the process up to 
trivial qubit chain (trunk). The methods can also be used for effective simulation of some quantum circuits corresponding 
to the binary tree structure. The modeling of more general qubit trees, as well as the relationship with the mapping used 
in the Bravyi--Kitaev transformation, are also briefly discussed.
\end{abstract}

\maketitle

\section{Introduction}
\label{Sec:Intro}

In earlier work \cite{Vla18} about effective modeling of quantum state transfer in qubit chains 
a question  about a problem to generalize suggested approach to arbitrary graphs was raised.
The presented work provides an extension of some methods used for qubit chains in \citeref{Vla18} to qubit 
trees together with appropriate applications.
It is also interesting from point of view of generalizations of Jordan--Wigner transformations \cite{JW}
to trees and more general graphs discussed in other works \cite{JWtr,JWlat,JWgr,SW18N,SW18}.

The approach developed in this work associates representations of Clifford algebras
and Spin group with ternary and binary qubit trees. 
It can be more naturally defined by ternary trees with transition to binary trees using some `pruning'. 
The application of similar ternary trees for fermion-to-qubit mapping was 
also discussed recently in \citeref{JKMN}.

Some preliminaries about Clifford algebras, Spin groups with application to construction of quantum gates 
are introduced in \Sec{Prel}. Representations of Clifford algebras and Spin groups using ternary qubit trees 
and deterministic finite automata are defined in \Sec{Tern} together with description of a `pruning process,'
{\em i.e.}, producing new trees by deleting of the branches.
The procedure can be also used for construction of binary qubit trees introduced in \Sec{Bin}. 
The binary trees can be 
considered as more natural generalization of some methods touched upon earlier in \citeref{Vla18} due to 
possibility to use some supplementary tools such as annihilation and creation operators discussed 
in \Sec{Ann}. 
The applications of the binary qubit trees to constructions of effectively modeled quantum circuits 
are outlined in \Sec{Eff} with some examples appropriate both for theory of quantum 
computations and communications.

The different scheme of qubit encoding by so-called Fenwick trees was also discussed
in \citeref{HTW17} for applications to Bravyi--Kitaev transformation \cite{BK00}. 
For trees of arbitrary size the number of children for some qubit nodes in such a case 
may not be limited. 
Such models can be encoded by an alternative version of binary trees outlined for completeness 
in \Sec{altbin} together with example of application to Bravyi--Kitaev encoding in \Sec{BK}.

\section{Preliminaries}
\label{Sec:Prel}

Let us recollect standard properties and definitions for Clifford algebras and Spin groups 
\cite{ClDir,Port} necessary in next sections.
For the vector space $\VV=\fld^n$ (where $\fld$ is $\RR$ or $\CC$) 
the Clifford algebra $\Clf(\VV)$ provides linear embedding of vector $\ve v \in \VV$
with property
\begin{equation}\label{sq}
 \e \colon \VV \longrightarrow \Clf(\VV),\quad
  \bigl(\e(\ve v)\bigr)^2 = -|\ve v|^2 \Id,
\end{equation}
where $\Id$ is the unit of the algebra and $|\ve v|$ is a norm of the vector. 
For a vector $\ve v \in \VV$ with coordinates $v_k$ the embedding is written
\begin{equation}\label{emb}
  \ve v = (v_1,\ldots,v_n), \quad	
  \e(\ve v) = \sum_{k=1}^n v_k \e_k,
\end{equation}
where $\e_k$ are {\em generators of Clifford algebra}. The possibility to 
work with complex vector spaces $\VV=\CC^n$ is desirable for many models below, but
some definitions and examples may be more naturally introduced for real case $\VV=\RR^n$.
The Minkowski (pseudo-Euclidean) norm is not considered here and for Euclidean case 
\Eq{sq} can be rewritten using \Eq{emb}
\begin{equation}\label{ClAR}
\{\e_j,\e_k\} \doteq \e_j \e_k + \e_k \e_j = -2 \delta_{jk}\Id,\quad j,k=1,\ldots,n.
\end{equation}

Due to relations \Eq{ClAR} maximal number of different  products of generators 
up to sign is $2^n$ and Clifford algebra with such dimension 
is called {\em universal} and denoted further $\UCl(n,\fld)$.
The natural non-universal examples are algebra of Pauli matrices
\begin{equation}
\sgm^x = \mat{0&1\\1&0},\quad
\sgm^y = \mat{0&-\mi\\ \mi&0},\quad
\sgm^z = \mat{1&0\\0&-1}
\label{PauliMat}
\end{equation} 
for $\VV=\CC^3$ and the {\em algebra of quaternions} $\Quat$ for 3D real space $\VV=\RR^3$. 
The dimension of such algebras is not maximal and one generator in such a case could be dropped 
to satisfy universality condition, but it may be not always justified due to
structure of a model.

For complex vector space with even dimension $\CC^{2m}$
the universal Clifford algebra $\UCl(2m,\CC)$ may be represented as 
$2^m \times 2^m$ complex matrix algebra \cite{ClDir}. 
The generators of $\UCl(2m,\CC)$ can be expressed using  
so-called Jordan--Wigner \cite{JW} representation 
\begin{equation}\label{Clgen}
\begin{array}{rcl}
\e_{2k-1} & = &
\mi\,{\underbrace{\sgm^z\otimes\cdots\otimes \sgm^z}_{k-1}\,}\otimes
\sgm^x\otimes\underbrace{\Id\otimes\cdots\otimes\Id}_{m-k}, \\
\e_{2k} & = &
\mi\,{\underbrace{\sgm^z\otimes\cdots\otimes \sgm^z}_{k-1}\,}\otimes
\sgm^y\otimes\underbrace{\Id\otimes\cdots\otimes\Id}_{m-k}, 
\end{array}
\end{equation}
where $k=1,\ldots,m$.

In odd dimensions the universal Clifford algebra $\UCl(2m+1,\CC)$ 
can be represented using block diagonal matrices  
\begin{equation}\label{ClBl}
\left(
\begin{array}{cc}
\mathbf A & \mathbf 0 \\
\mathbf 0 & \mathbf B
\end{array}
\right) \in \UCl(2 m+1,\CC), 
\quad \mathbf A, \mathbf B \in \UCl(2 m,\CC),
\end{equation} 
{\em i.e.}, as the direct sum of {\em two} $\UCl(2m,\CC)$,
but an irreducible representation with the half of maximal dimension 
also exists. It may be treated as $\UCl(2m,\CC)$ with the additional generator that can be expressed 
up to possible imaginary unit multiplier as product of all $2m$ generators. 
For representation \Eq{Clgen} it may be written
\begin{equation}
\label{Clgen'}
\e_{2m+1} = \mi\,\underbrace{\sgm^z\otimes\cdots\otimes \sgm^z}_{m}\,.
\tag{\ref{Clgen}$'$}
\end{equation}

Such a case is essential for many examples considered below.
Using $2m$ generators \Eq{Clgen} together with the extra one \Eq{Clgen'} denoted for certainty as
$\e_j^{(2m)}$ the representation of generators $\e_j^{(2m+1)}$ respecting \Eq{ClBl}
for universal Clifford algebra $\UCl(2m+1,\CC)$  can be written as 
\begin{equation}\label{Clgen1}
\e^{(2m+1)}_j=\sgm^z\otimes\e_j^{(2m)}, \quad j=1,\ldots,2m+1.
\end{equation}

The group Spin$(n)$ is defined as a subset of $\UCl(\RR,n)$ generated by all possible products of {\em even}
number of elements $\e(\ve v)$ with different vectors $\ve v$ of unit length
\begin{equation}\label{Spgr}
\begin{split}
 \sps= \e(\ve v_1) \e(\ve v_2) \cdots \e(\ve v_{2k}),\quad&
   \ve v_j \in \RR^n,\\ |\ve v_j|=1, \quad &j=1,\ldots,2k.
\end{split}   
\end{equation}
The basic property of Spin$(n)$ is expression of orthogonal group as
\begin{equation}\label{SpOr}
 \sps \, \e(\ve v) \, \sps^{-1} = \e(\ve v'),\quad \ve v' =\mt R_{\sps} \ve v, \quad \mt R_{\sps} \in \text{SO}(n),
\end{equation}
{\em i.e.}, $\mt R_{\sps}$ is some $n$-dimensional rotation.
It should be noted, that {\em two} elements $\pm \sps \in \text{Spin}(n)$ in \Eq{SpOr} correspond to 
{\em the same} transformation \mbox{$\mt R_{\sps} \in \text{SO}(n)$}.
Thus, Spin$(n)$ group {\em doubly covers} SO$(n)$. 

The Spin group also can be described as the Lie group.
The universal Clifford algebra $\UCl_n=\UCl(\fld,n)$ is a Lie algebra with respect to
the bracket operation
$$[\cla,\clb]= \cla\clb-\clb\cla, \quad \cla,\clb \in \UCl_n.$$
For the Lie group Spin$(n)$ the Lie algebra spin$(n)$ is a subalgebra 
of $\UCl_n$ with the basis $\e_j\e_k$, $1\le j<k \le n$.
The Lie algebra so$(n)$ of the orthogonal group is isomorphic with
spin$(n)$.

The representation of Spin$(n)$ groups using the Clifford algebras discussed above
has dimension $2^n$, but the both spin$(n)$ and so$(n)$ have dimension only $n(n-1)/2$.
The Lie algebraic approach is also important due to direct relation with
Hamiltonians of quantum gates \cite{Vla18,Vla0}. 

There is some subtlety, because exponential map producing an element of 
the Spin group is $\spn A_\epsilon = \exp(\epsilon \cla)$, but 
in the physical applications expressions with the generators are often 
written with imaginary unit multiplier, {\em e.g.}, the  quantum gates near identity should be written \cite{DV95}
\begin{equation}\label{deltU}
 \delta \op U = e^{\mi \epsilon \op H} \simeq \Id + \mi\epsilon \op H,
 \quad \epsilon \rightarrow 0 .
\end{equation}
In such a case the imaginary unit should also appears in anticommutators.
For example, the commutator algebra with the bracket operation $\mi[\op H_a,\op H_b]$
appears in a proof of two-qubit gates universality \cite{DV95}.
The set of gates represented in such a way is universal if
elements $\op H$ from \Eq{deltU} generate entire Lie algebra of unitary
group by the commutators.

Similar Lie-algebraic approach with Clifford algebras can be used both
for construction of universal and non-universal sets of two-qubit gates \cite{Vla0}.
The basis of the Lie algebra spin$(2m)$ consists of quadratic elements $\e_j\e_k$.
The construction of the Lie algebra spin$(2m)$ using \Eq{Clgen} represents
the Spin$(2m)$ group as some subgroup of the unitary group U$(2^m)$.
 
Let us consider four consequent generators $\e_{2k-1}$, $\e_{2k}$, $\e_{2k+1}$, $\e_{2k+2}$. 
The linear combinations of six different quadratic elements produced from the generators for 
particular representation \Eq{Clgen} correspond to Hamiltonians of some one- and two-qubit gates.
For different $k$ it generates the non-universal set of quantum gates on nearest-neighbor qubits
often called {\em matchgates} \cite{Val1,TD2}. 

The Jordan--Wigner representation of generators for Clifford algebra \Eq{Clgen}
is not unique. Alternative methods based on tree-like structures are discussed in next sections.

\section{Ternary Trees}
\label{Sec:Tern}

Let us consider nine generators
\begin{equation}\label{tgen9}
{\renewcommand{\arraystretch}{0.9}
	\begin{array}{rcl}
	\ttg_1 & = & \mi\sgm^x \otimes \sgm^x \otimes \Id \otimes \Id, \\
	\ttg_2 & = & \mi\sgm^x \otimes \sgm^y \otimes \Id \otimes \Id, \\
	\ttg_3 & = & \mi\sgm^x \otimes \sgm^z \otimes \Id \otimes \Id, \\
	\ttg_4 & = & \mi\sgm^y \otimes \Id \otimes \sgm^x \otimes \Id, \\
	\ttg_5 & = & \mi\sgm^y \otimes \Id \otimes \sgm^y \otimes \Id, \\
	\ttg_6 & = & \mi\sgm^y \otimes \Id \otimes \sgm^z \otimes \Id, \\
	\ttg_7 & = & \mi\sgm^z \otimes \Id \otimes \Id \otimes \sgm^x, \\
	\ttg_8 & = & \mi\sgm^z \otimes \Id \otimes \Id \otimes \sgm^y, \\
	\ttg_9 & = & \mi\sgm^z \otimes \Id \otimes \Id \otimes \sgm^z.
	\end{array}}
\end{equation}

A shorter notation is often used further, {\em e.g.},
\begin{equation}
{\renewcommand{\arraystretch}{1.2}
\begin{array}{lll}
\ttg_1 = \mi\sgm^x_1 \sgm^x_2, & 
\ttg_2 = \mi\sgm^x_1 \sgm^y_2, &  
\ttg_3 = \mi\sgm^x_1 \sgm^z_2, \\
\ttg_4 = \mi\sgm^y_1 \sgm^x_3, & 
\ttg_5 = \mi\sgm^y_1 \sgm^y_3, &  
\ttg_6 = \mi\sgm^y_1 \sgm^z_3, \\ 
\ttg_7 = \mi\sgm^z_1 \sgm^x_4, & 
\ttg_8 = \mi\sgm^z_1 \sgm^y_4, &  
\ttg_9 = \mi\sgm^z_1 \sgm^z_4, 
\end{array}}
\tag{\ref{tgen9}$'$}\label{tgen9'}
\end{equation}
where $\sgm^\mu_j$ denotes Pauli matrix $\mu =x,y,z$ acting on qubit with index $j$. 

The universal Clifford algebra could be defined using eight generators instead of nine and
product of all $\ttg_k$ is identity up to possible multiplier with some power of imaginary 
unit denoted further as
\begin{equation}
\pmi \in \{\pm1,\pm\mi\},\quad \pmi^4 = 1.
\label{idmult}
\end{equation}

\begin{figure}[htb]
	\begin{center}	
		\includegraphics[scale=0.75]{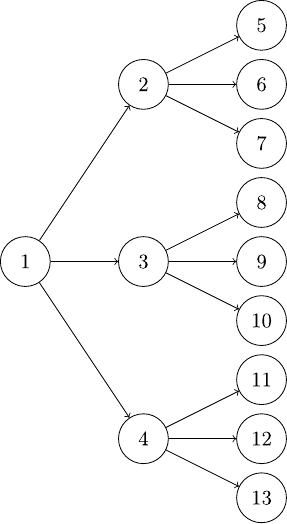}
	\end{center}
	\caption{Ternary $\CFtr\Lvt$-tree with $\Lvt=3$ }
	\label{Fig:tertree}
\end{figure}

\begin{figure}[htb]
\begin{center}	
\includegraphics[scale=0.75]{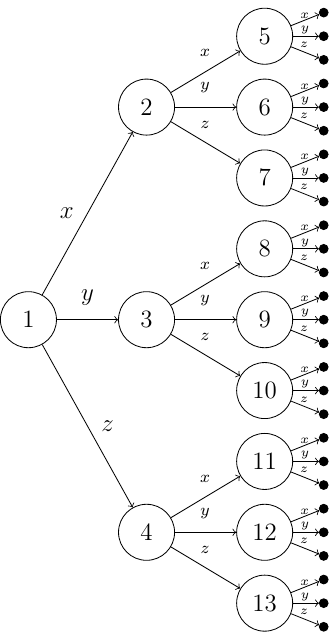}		
\end{center}
\caption{DFA from ternary $\CFtr\Lvt$-tree {\em extended} by leaf nodes}
\label{Fig:terDFA}
\end{figure}

Nine generators \Eq{tgen9} demonstrate natural threefold symmetries derived from Pauli matrices. 
The generalization for arbitrary power of three using ternary trees is discussed below. 
For the initial example \Eq{tgen9} it corresponds to four qubits 
nodes $j=1,\ldots,4$ represented by lower indexes in \Eq{tgen9'} with root is $j=1$ and three child 
nodes \mbox{$j = 2,3,4$} are associated with three generators each. 
Such construction can be generalized, {\em e.g.}, similar example with tree for {\em thirteen qubits} 
is provided below (see \Fig{tertree}) with scheme of {\em twenty seven generators} 
is depicted on \Fig{terDFA}.

Let us recollect some properties of rooted trees \cite{Kn1,Disc}.
A node of $n$-ary tree has up to $n$ children, the nodes without
any child are called {\em terminal} nodes or {\em leafs}. 
The {\em level} $\ell$ is defined here as the number of nodes in the path 
from the root.
The maximal level of nodes in a tree is denoted further as $\Lvt$ and, thus, the {\em height} of the tree is $\Lvt-1$. 

Ternary or binary trees with maximal number of nodes for given $\Lvt$ 
are denoted here for certainty as `$\CFtr\Lvt$-trees'. 
It could be formally described
using definitions from \citeref{Disc} as {\em directed rooted complete full ternary 
(or binary) tree with height $\Lvt-1$}.
An auxiliary root with index zero can be also attached in some constructions
below to the first node producing trees of height $\Lvt$. 
Such a method is relevant to \Eq{genClpl} and \Eq{genClpp} below.
It is also used for producing of $\oCFtr{\Lvt}$-tree from $\CFtr\Lvt$-tree in \Sec{altbin}.

\smallskip

Number of nodes in a {\em ternary} $\CFtr{\Lvt}$-tree is
\begin{equation}\label{nt}
 m_\Lvt = \sum_{k=0}^{\Lvt-1} 3^k = \frac{3^\Lvt-1}{2}.
\end{equation}
Let us start with three generators $\ttg_1^{(3)} = \mi\sgm^x$, $\ttg_2^{(3)} = \mi\sgm^y$, $\ttg_3^{(3)} = \mi\sgm^z$
for $\Lvt=1$.
For any $\Lvt>1$, $3^{\Lvt+1}$ anticommuting generators for ternary $\CFtr{\Lvt+1}$-tree can be produced 
by recursion $\Lvt\to\Lvt+1$  using $3^\Lvt$ anticommuting generators defined for $\CFtr{\Lvt}$-tree 
\begin{eqnarray}
\ttg_{3j-2}^{(3^{\Lvt+1})} &=& \ttg_j^{(3^\Lvt)} \otimes {\underbrace{\Id\otimes\cdots\otimes\Id}_{j-1}\,}%
 \otimes \sgm^x \otimes \underbrace{\Id\otimes\cdots\otimes\Id}_{3^\Lvt-j}\, , \notag\\
\ttg_{3j-1}^{(3^{\Lvt+1})} &=& \ttg_j^{(3^\Lvt)} \otimes {\underbrace{\Id\otimes\cdots\otimes\Id}_{j-1}\,}%
 \otimes \sgm^y \otimes \underbrace{\Id\otimes\cdots\otimes\Id}_{3^\Lvt-j}\, ,\label{tgenrec} \\ 
\ttg_{3j}^{(3^{\Lvt+1})} &=& \ttg_j^{(3^\Lvt)} \otimes {\underbrace{\Id\otimes\cdots\otimes\Id}_{j-1}\,}%
\otimes \sgm^z \otimes \underbrace{\Id\otimes\cdots\otimes\Id}_{3^\Lvt-j}, \notag 
\end{eqnarray} 
where $j = 1,\ldots,3^\Lvt$ and the total number of terms in the tensor product is \mbox{$m_{\Lvt+1} = m_{\Lvt}+3^\Lvt$}.
All generators in \Eq{tgenrec} anticommute --- in different triples due to 
terms $\ttg_j^{(3^\Lvt)} $ and in the same triple due to terms $\sgm_j^\mu$ ($\mu =x,y,z$).

Let us prove recursively that any $3^\Lvt-1$ generators between $\ttg_j^{(3^\Lvt)}$ 
generate whole basis for universal Clifford algebra $\UCl(2m_{\Lvt},\CC)$. 
Let us start with useful property: the product of all $3^\Lvt$ generators is $\pmi\Id$. 
It is true for $\Lvt=1$, $\ttg_k^{(3)}$, $k=1,2,3$ and for any $\Lvt+1$ 
it is derived directly from \Eq{tgenrec}. Due to such property any chosen generator up to $\pmi$ multiplier 
is represented as product of all other generators and can be dropped. Thus, any
$3^\Lvt-1$ generators between $3^\Lvt$ can be used as a basis of $\UCl(2m_{\Lvt},\CC)$.

The standard basis of $\UCl(2m_{\Lvt},\CC)$ is naturally expressed as $4^{m_{\Lvt}}$ tensor products
using {\em Pauli basis}, {\em i.e.}, three Pauli matrices and $2 \times 2$ unit matrix. 
Let us show, that the basis can be also represented (not necessary in unique way) 
by products of $\ttg^{(3^\Lvt)}_k$. It is again true for $\Lvt=1$ and $\UCl(2,\CC)$. 
Let us consider $\Lvt+1$ for some $\Lvt \ge 1$ with the basis of $\UCl(2 m_{\Lvt},\CC)$  
expressed by products of $\ttg^{(3^\Lvt)}_k$.
Arbitrary basic element $\clb$ of $\UCl(2 m_{\Lvt+1},\CC)$ can be represented as tensor products with
$m_{\Lvt+1}$ elements of Pauli basis. The product of three generators for any $j$ in \Eq{tgenrec} is
$$\pmi\,\ttg_j^{(3^\Lvt)} \otimes {\underbrace{\Id\otimes\cdots\otimes\Id}_{3^\Lvt}},$$ 
so, the first $m_\Lvt$ terms in $\clb$ can be rewritten by product 
of such triples due to previous steps of recursion. 
Three possible products of two generators with given $j$ in \Eq{tgenrec} are
$$\pmi\,{\underbrace{\Id\otimes\cdots\otimes\Id}_{m_{\Lvt}+j-1}\,}%
\otimes \sgm^\mu \otimes \underbrace{\Id\otimes\cdots\otimes\Id}_{3^\Lvt-j}, \quad \mu =x,y,z,$$ 
and remaining last $3^\Lvt$ terms of $\clb$ can be also expressed using 
products of such pairs. 
So, any element $\clb$ of standard basis $\UCl(2 m_{\Lvt+1},\CC)$ with $m_{\Lvt+1}=m_{\Lvt}+3^\Lvt$ terms
is some product of $\ttg_k^{(3^{\Lvt+1})}$. 

It was also shown, that any element can be expressed up to $\pmi$ as product of other generators.
In such a case the construction with one dropped element corresponds to {\em universal} Clifford 
algebra. \qed

Each generator $\ttg^{(3^\Lvt)}_k$, $k=1,\ldots,3^\Lvt$ has $m_\Lvt{=}(3^\Lvt-1)/2$ terms in tensor product 
with only $\Lvt$  (non-unit) Pauli matrices, because recursion \Eq{tgenrec} appends only one non-unit term 
for each level. 
The scheme of such terms may be represented by directed ternary $\CFtr{\Lvt}$-tree 
with first qubit as root, see \Fig{tertree}. 
Each triple of generators in \Eq{tgenrec} formally corresponds to path from the root of the tree
to {\em leaf} nodes.

For example, the tree with three levels represented on \Fig{tertree} may illustrate structure 
of nine triples with twenty seven generators: 
$\ttg_1^{(27)} = \mi\sgm^x_1\sgm^x_2\sgm^x_5$, 
\mbox{$\ttg_2^{(27)} = \mi\sgm^x_1\sgm^x_2\sgm^y_5$}, 
$\ttg_3^{(27)} = \mi\sgm^x_1\sgm^x_2\sgm^z_5$,
$\ttg_4^{(27)} = \mi\sgm^x_1\sgm^x_2\sgm^x_6$, $\ldots,$
$\ttg_{27}^{(27)} = \mi\sgm^z_1\sgm^z_4\sgm^z_{13}$. 

The representation with tree provides yet another explanation of anticommutativity
of all $\ttg_j^{(m_\Lvt)}$. Any two `branches' of tree have some common part corresponding 
to qubits with the same index and non-unit tensor factors, but only last pair of Pauli 
matrices in common subsequences (corresponding to `fork node' for pair of branches) may differ. 
Such approach produces an illustrative argument for the generalization with arbitrary ternary trees.

Let us first extend the model to provide formal definition using some methods 
from theory of {\em deterministic finite automata} (DFA) \cite{Con71,InCom}. 
The model of deterministic finite automaton below uses {\em extension} \cite{Kn1} of ternary $\CFtr{\Lvt}$-tree 
with basic nodes representing qubits and three additional {\em output nodes} for 
each leaf. For more general ternary trees discussed further number of children 
for any qubit node is added up to three by new output nodes.

Each link is marked by letters $x,y,z$ representing possible transition 
between nodes, see \Fig{terDFA}. The {\em word} (sequence of letters
$x,y,z$) corresponding to path from the root to output nodes is {\em recognized}
by deterministic finite automaton. The sequence of nodes generated by such transition 
represents generator expressed as product of terms with Pauli matrices indexed by number of node and letter, 
{\em e.g.},  
\mbox{$xxx \to \sgm_1^x\sgm_2^x\sgm_5^x$, $\ldots$,}
\mbox{$xyz \to \sgm_1^x\sgm_2^y\sgm_6^z$, $\ldots$,}
\mbox{$zyx \to \sgm_1^z\sgm_4^y\sgm_{12}^x$, $\ldots$},
$zzz \to \sgm_1^z\sgm_4^z\sgm_{13}^z$ for \Fig{terDFA}.

More generally, if some sequence $\mu_1 \mu_2 \ldots \mu_\ell$ of letters $\mu_k \in \{x,y,z\}$ for
$k = 1,\ldots,\ell$ is recognized by deterministic finite automaton and generates sequence of nodes (path)
\begin{equation}\label{DFApath}
j_1 \xrightarrow{\mu_1} j_2 \xrightarrow{\mu_2} 
\cdots \xrightarrow{\mu_{\ell-1}} j_\ell \xrightarrow{\mu_l} o_{\ell+1}
\end{equation} 
with root $j_1 = 1$ and $o_{\ell+1}$ is the output node, the generator is
\begin{equation}\label{tauDFA}
\ttg_{o_{\ell+1}}=\mi\sgm_{j_1}^{\mu_1}\sgm_{j_2}^{\mu_2}\cdots\sgm_{j_\ell}^{\mu_\ell} = \mi\prod_{k=1}^\ell\sgm_{j_k}^{\mu_k}.
\end{equation}

\medskip

The model with deterministic finite automaton and \Eq{tauDFA} can be applied for a general ternary tree
for a level $\ell$ that is not necessary equal to the maximal $\Lvt$ and the number 
of outbound links for each node
may be from zero to three. Let us start with a ternary $\CFtr{\Lvt}$-tree 
discussed above with maximal number of qubit nodes $m_{\rm q}=(3^\Lvt-1)/2$ 
and $n_{\rm g}=3^\Lvt$ anticommuting generators
\begin{equation}\label{nqng}
 n_{\rm g} = 2m_{\rm q}+1.
\end{equation}
The \Eq{nqng} is also valid for any subtree.

Other ternary trees can be produced by recursive process of `pruning' discussed below.
Let us delete all nodes and generators of subtree $\sbtr$ originated from node 
$j_{\sbtr}$ attached to parent node $j_p$ by link with label $\mu_p \in \{x,y,z\}$.
Let us also add the new element including only initial common sequence of nodes in products \Eq{tauDFA} 
coinciding for all deleted nodes of the subtree $\sbtr$
\begin{equation}\label{taucut}
 \ttg_{\sbtr} = \mi\sgm_1^{\mu_1} \cdots \sgm_{j_p}^{\mu_p}.
\end{equation}

The tree and all its subtrees after any deletion also meet \Eq{nqng},
because
$$
 n'_{\rm g} = n_{\rm g} - n^{\sbtr}_{\rm g} + 1 = (2m_{\rm q}+1)-(2m^{\sbtr}_{\rm q}+1)+1=2m'_{\rm q}+1,
$$
where $n'_{\rm g}$, $m'_{\rm q}$ and $n^{\sbtr}_{\rm g}$, $m^{\sbtr}_{\rm q}$ denote 
parameters (number of generators, number of qubit nodes) for
produced tree and deleted subtree respectively.

The new element \Eq{taucut} anticommutes with all elements except deleted. Let us also prove
that product of $n'_{\rm g}$ generators for new tree is $\pmi\Id$, 
there $\pmi$ is possible unessential multiplier \Eq{idmult}. 
For initial ternary $\CFtr{\Lvt}$-tree \Eq{nqng} is true and the product of all generators 
was already calculated earlier. Any subtree of the $\CFtr{\Lvt}$-tree is also ternary $\CFtr{\Lvt'}$-tree
for some $\Lvt' < \Lvt$ and product of all generators for such subtree is 
$$
 \prod_{k \in {\sbtr}}\ttg_k = (\ttg_{\sbtr})^{n_{\rm g}^{\sbtr}}\,(\pm\Id) 
 = \mp\ttg_{\sbtr},
$$
because $n_{\rm g}^{\sbtr}$ is odd and $(\ttg_{\sbtr})^2=-1$. So, after each deletion the products of
all generators of {\em deleted trees} up to sign are equal with corresponding $\ttg_{\sbtr}$
and total product of {\em all elements} is always $\pmi\Id$.

Let us prove, that for any tree with $m'_{\rm q}$ qubit nodes obtained by such pruning,
the products of any subset with $n_{\rm g}'- 1 = 2m'_{\rm q}$ generators may be used as a basis of
universal Clifford algebra $\UCl(2m'_{\rm q},\CC)$. Let us again for simplicity start 
with all $n_{\rm g}' = 2m'_{\rm q}+1$ generators, because any generator may be expressed as 
product of other generators.

Let us note, that each deletion in process of pruning may be treated also as two stage process: 
(1) to drop multipliers with Pauli matrices for excluded qubit nodes from all products and 
(2) to remove duplicates from list of generators. The approach is also correct for
description of whole pruning as a series of consequent deletions.

Let us consider final tree as subtree of ternary $\CFtr{\Lvt}$-tree.
Any element of standard basis of the Clifford algebra for qubits from 
this subtree can be represented by product of generators of initial tree.
If to drop Pauli matrices for extra qubits from generators in such 
products the result may only change sign, but now it includes only terms
those equal with generators of subtree. 
Thus, the terms is a basis of Clifford algebra for the final tree.
\qed

\medskip

Let us describe formal procedure for construction of generators from 
arbitrary ternary tree produced by the pruning described above:
\begin{itemize}
\item Ternary tree should be extended by adding of {\em terminal} (output) nodes, 
 {\em i.e.}, all initial nodes with number of children $n_{\rm c} < 3$ 
 should be connected with $3-n_{\rm c}$ new leafs associated with generators.
\item Now all non-terminal (initial) nodes have three output links 
marked by triple of labels $x,y,z$. Such a tree also may be 
 considered as a deterministic finite automaton. 
\item Any path from root to terminal node is described by
 analogue of \Eq{DFApath} with $l$ is level of the node and
 the generator for each terminal node can be expressed as \Eq{tauDFA}. 
\item Formally, a possible sequence of letters $\mu_k \in \{x,y,z\}$ in \Eq{DFApath} 
 corresponds to {\em a word recognized by the deterministic finite automaton} and any generator is
 represented in such a way by product of Pauli matrices \Eq{tauDFA}.
\end{itemize}

Let us summarize construction of generators using {\em extended ternary tree}. Rooted
directed ternary tree is defined by set of qubit nodes $j=1,\ldots,m$
and directed links between pairs of nodes. Any node except root has one parent 
and up to three children.
The links are marked by labels $x,y,z$. 

 Let us first for any qubit
node $j$ define an auxiliary operator ({\em stub}) $\stub_j$. 
For root node $j=1$,  $\stub_1 = \mi\Id$ and for any 
child node $k$ linked with a parent node $j$ by link with a label
$\mu \in \{x,y,z\}$
\begin{equation}\label{stub}
 j \xrightarrow{\mu} k :\quad
 \stub_k = \stub_j \sgm_j^\mu.
\end{equation}

Now for any node $j$ with less than three children $n_{\rm c}$ it is necessary to attach
$n_o=3-n_{\rm c}$ {\em output} generator nodes with appropriate unique indexes $\gj$ by new links 
for missing labels \mbox{$\mu \in \{x,y,z\}$}.

The maximal total number of outbound links for $m$ nodes is $3m$, but
$m-1$ children are qubits nodes (because all of them except root have one parent). 
Thus, number of generator nodes satisfies \Eq{nqng}
$$ n_{\rm g} = 3m-(m-1) = 2m+1.$$

The generator associated with each such node is defined as
\begin{equation}\label{stubgen}
\begin{split}
\ttg_\gj  = \ttg_{j\sep\mu} = \stub_j \sgm_j^\mu, \quad
 &\gj = 1,\ldots,2m+1,\\ &j=1,\ldots,m.
\end{split} 
\end{equation}

An alternative notation $\ttg_{j\sep\mu}$ is introduced for convenience 
in \Eq{stubgen}. Any generator may be expressed in such a way $\ttg_\gj = \ttg_{j\sep\mu}$
after choosing of some map to set of consequent indexes $\gj = \gj(j,\mu)$, 
but number of elements $\ttg_{j\sep\mu}$ is bigger, $3m > 2m+1$.
Redundant $\ttg_{j\sep\mu}$ correspond to products of generators denoted earlier as
$\ttg_{\sbtr}$ \Eq{taucut}. \qed

\medskip

The \Eq{stubgen} together with definition of stub operator \Eq{stub} formalizes 
\Eq{tauDFA} used earlier without necessity to introduce an enveloping $\CFtr{\Lvt}$-tree. 

\smallskip

For the ternary $\CFtr{\Lvt}$-tree deterministic finite automaton recognizes 
any sequences with $\Lvt$ letters and resulting $3^\Lvt$ generators are
attached to leafs of qubit tree \Fig{terDFA}. Number of nodes for such a tree
is $(3^\Lvt-1)/2$ \Eq{nt}.

For more general ternary tree with $m$ nodes produced with the method discussed above 
the number of generator leafs (DFA output nodes) on the extended tree is always $2 m+1$. The product of
all generators is proportional to identity. It was already discussed that
any subset with $2 m$ generators may be used for construction of 
universal Clifford algebra $\UCl(2 m,\CC)$.

Let us consider yet another formal construction of $\UCl(2 m+1,\CC)$ without necessity to
get rid of one generator. Let us introduce an auxiliary node with index zero to extend
the set of generators to $m+1$ qubits using straightforward method, {\em cf\/} \Eq{Clgen1}
\begin{equation}\label{genClpl}
  \acute{\e}_j = \sgm^z \otimes \ttg_j,\quad j=1,\ldots,2m+1.
\end{equation}   
The products of $2m+1$ elements \Eq{genClpl} is $\sgm_0^z$ 
and, thus, $\UCl(2 m+1,\CC)$ can be generated by \Eq{genClpl} 
using standard representation with block diagonal matrices,
see \Eq{ClBl}.

The {\em even subalgebra} $\UCl_0$ is generated by products of even number of generators $\acute{\e}_j$ 
\Eq{genClpl}. The cancellation of $\sgm_0^z$ in products illustrates natural isomorphism
$$\UCl_0(2m+1,\CC) \simeq \UCl(2 m,\CC)$$
and it also produces representation of Spin$(2m+1)$ group by all $2m+1$ 
elements $\ttg_j \in \UCl(2 m,\CC)$.

\smallskip

For $m > 1$ the Spin$(2m+2)$ can be also represented in a similar way. Let us consider
construction of Spin groups as Lie algebras \cite{ClDir} recollected in \Sec{Prel}. 
In such a case the element may be expressed as exponent of linear 
combinations of quadratic terms $\e_j\e_k$. 

Let us again introduce an extra zero node, but for alternative representation 
of $2m+2$ generators instead of \Eq{genClpl} should be used
\begin{equation}\label{genClpp}
\begin{array}{ll}
\grave{\e}_j = \sgm^x \otimes \ttg_j,& j=1,\ldots,2m+1,\\
\grave{\e}_0 = \sgm^y \otimes \Id \otimes \cdots  \otimes \Id.
\end{array}
\end{equation} 
The products of two such elements are either 
$\Id \otimes (\ttg_j\ttg_k)$ or $\sgm^z \otimes \ttg_l$, 
where $j,k,l=1,\ldots,2m+1$. 
The quadratic terms can be expressed as block-diagonal matrices \Eq{ClBl}. 
For $m>1$ all $\ttg_j\ttg_k$ with $j<k$ and $\ttg_l$ are different 
and exponents of matrices with linear combination of such elements 
$\exp(\mathbf A) \in \UCl(2 m,\CC)$ can be used for construction of 
irreducible representation of Spin$(2m+2)$. 
It is not true for $m=1$ due to $\ttg_1\ttg_2=\ttg_3$, {\em e.g.}, 
for quaternions or Pauli matrices $\sgm_x\sgm_y=\mi\sgm_z$.

\medskip

A standard representation of Clifford algebra may be considered as an extreme
case of pruning into a chain of $z$-linked nodes. 
At least two generators ($x$, $y$) are attached to each node with an 
additional one ($z$) on the end.
Such a degenerate tree corresponds to $2m$ Jordan--Wigner generators \Eq{Clgen}
$$
\begin{array}{lcr}
 \e_{2k-1} &=& \mi\sgm_1^z\cdots\sgm_{k-1}^z\,\sgm_k^x\\
 \e_{2k} &=& \mi\sgm_1^z\cdots\sgm_{k-1}^z\,\sgm_k^y
 \end{array}
$$
for $k=1,\dots,m$
together with \Eq{Clgen'} $$\e_{2m+1} = \mi\sgm_1^z\cdots\sgm_{2m}^z.$$

\section{Binary Trees}
\label{Sec:Bin}
 
\begin{figure}[htb]
	\begin{center}	
		\includegraphics[scale=0.75]{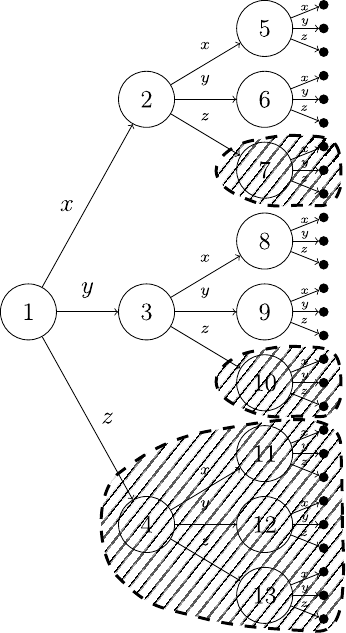}		
	\end{center}
	\caption{Binary ({\it x-y}) tree obtained from ternary on \Fig{terDFA}} 
	\label{Fig:ter2bin}
\end{figure}

Binary $\CFtr{\Lvt}$-trees can be introduced formally by deleting of all nodes attached 
to $z$-links of the ternary $\CFtr{\Lvt}$-trees, see \Fig{ter2bin}.
The term {\em binary x-y tree} may be also used sometimes to distinguish that from 
an alternative construction with deleted $y$-links, but such `{\it x-z} trees' 
are introduced only in \Sec{altbin}.

\begin{figure}[htb]
	\begin{center}	
		\includegraphics[scale=0.75]{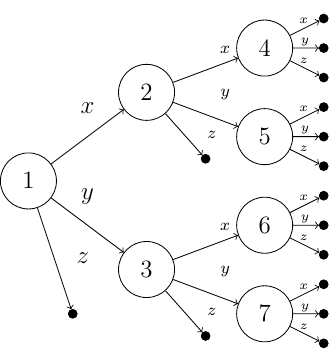}		
	\end{center}
	\caption{DFA for binary tree with additional leaf nodes}
	\label{Fig:binDFA}
\end{figure}

The deterministic finite automaton for such binary tree produces three generators 
for terminal qubit nodes with maximal level $l=\Lvt$, but only one generator 
for other qubit nodes with $l<\Lvt$, see \Fig{binDFA}.

\begin{figure}[htb]
	\begin{center}	
		\parbox[c]{4cm}{\includegraphics[scale=0.75]{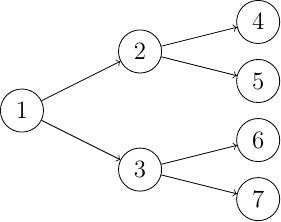}}
        \parbox[c]{5cm}{\includegraphics[scale=0.75]{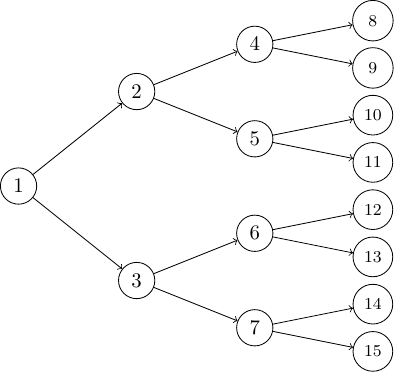}}
	\end{center}
	\caption{Binary $\CFtr{\Lvt}$-trees for $\Lvt=3$ and $\Lvt=4$}
	\label{Fig:bintree}
\end{figure}

The binary $\CFtr{\Lvt}$-tree has $2^\Lvt-1$ qubit nodes.
With `enumeration along levels' the nodes $j =1,\ldots, 2^{\Lvt-1}-1$ have 
two children $2j$ and $2j+1$, except leafs $j = 2^{\Lvt-1},\ldots,2^\Lvt-1$, 
see \Fig{bintree}. 

The {\em stub operator} $\stub_j$ \Eq{stub} used for construction of generators \Eq{stubgen}
can be constructed for binary case in the similar way as $\stub_1 = \mi\Id$ and
\begin{equation}\label{binstub}
 \stub_{2j} = \stub_j \sgm_j^x, \quad 
 \stub_{2j+1} = \stub_j \sgm_j^y.
\end{equation}

For binary tree with $m_{\rm q}=2^\Lvt-1$ qubits discussed earlier
the structure of generators is described by extension into {\em ternary} tree, see \Fig{binDFA}.
Qubits with indexes $j=1,\ldots,2^{\Lvt-1}-1$ have only one generator node, but three
generators are linked to remaining $2^{\Lvt-1}$ terminal qubit nodes $k=2^{\Lvt-1},\ldots,2^\Lvt-1$
with maximal level $\Lvt$. Thus, total number of generators meets \Eq{nqng}
$$n_{\rm g}=2^{\Lvt-1}-1+3\cdot 2^{\Lvt-1} = 2^{\Lvt+1}-1 = 2m_{\rm q}+1. $$
Here the `redundant' notation for generators used in \Eq{stubgen}
may be more illustrative
\begin{subequations}\label{bgen}
\begin{eqnarray}
 \ttg_{j\sep z} =\stub_j\sgm_j^z,&\quad &j=1,\ldots,2^{\Lvt-1}-1,\\
 \ttg_{j\sep\mu}=\stub_j\sgm_j^\mu ,&\quad &j=2^{\Lvt-1},\ldots,2^\Lvt-1, \notag\\
  &&\mu=x,y,z.   
\end{eqnarray}
\end{subequations}
For example of binary tree with $\Lvt=2$ and three qubits, seven generators
can be written as
\begin{equation}\label{bgen2}
\begin{split}
 \ttg_{1\sep z} =\mi\sgm_1^z,\quad
 \ttg_{2\sep\mu} &=\mi\sgm_1^x\sgm_2^\mu,\quad
 \ttg_{3\sep\mu} =\mi\sgm_1^y\sgm_3^\mu,
  \\ \mu&=x,y,z.
\end{split}  
\end{equation}

The sequence of terms with index $z$ from \Eq{bgen} can be also extended to all qubits.
Let us use notation $\btg_j$ or $\btg_j^{(n_{\rm g})}$, $j=1,\dots,n_{\rm g}=2^{\Lvt+1}-1$ for generators
with a consequent indexing with ranges
\begin{subequations}\label{bgind}
	\begin{eqnarray}
	&\btg_j^{(n_{\rm g})} = \ttg_{j\sep z},  &\quad  j=1,\ldots,2^\Lvt-1,\\
	&
	{\renewcommand{\arraystretch}{1.5}
	\left.	\begin{array}{r}
	\btg_{2j}^{(n_{\rm g})}\ =\ttg_{j\sep x}\\
    \btg_{2j+1}^{(n_{\rm g})}\ =\ttg_{j\sep y}  
	\end{array} \right\}
    }
	&\quad j=2^{\Lvt-1},\ldots,2^\Lvt-1.   
	\end{eqnarray}
\end{subequations}
Thus, for binary tree with three qubits \Eq{bgen2} can be rewritten
\begin{equation}\label{bgind2}
\begin{split}
\btg_1^{(7)} = \mi\sgm_1^z,\quad
&\btg_2^{(7)} = \mi\sgm_1^x\sgm_2^z,\quad
\btg_3^{(7)} = \mi\sgm_1^y\sgm_3^z,\\
&\btg_4^{(7)} = \mi\sgm_1^x\sgm_2^x,\quad
\btg_5^{(7)} = \mi\sgm_1^x\sgm_2^y,\\
&\btg_6^{(7)} = \mi\sgm_1^y\sgm_3^x,\quad
\btg_7^{(7)} = \mi\sgm_1^y\sgm_3^y.
\end{split}
\end{equation}

The indexing \Eq{bgind} is convenient due to properties of
triples with generators $\btg_j$, $\btg_{2j}$, $\btg_{2j+1}$.
Let us denote
\begin{equation}\label{qgen}
\begin{split}
 \btq^x_j = \mi\btg_{2j+1}\btg_j,\quad
 &\btq^y_j = \mi\btg_j\btg_{2j},\quad 
 \btq^z_j = \mi\btg_{2j}\btg_{2j+1}, \\ &j=1,\ldots,2^\Lvt-1. 
\end{split} 
\end{equation}
The terms \Eq{qgen} are trivial for index $j$ corresponding to terminal qubit nodes 
with three generators 
\begin{equation}\label{qgeno}
 \btq^\mu_j = \sgm^\mu_j,\quad j=2^{\Lvt-1},\ldots,2^\Lvt-1, \quad \mu=x,y,z.
\end{equation}
For nodes with single generator first pair of expressions \Eq{qgen}
can be associated with links of binary tree
\begin{equation}\label{qgenxy}
\btq^y_j = \sgm^y_j\sgm^z_{2j},\quad 
\btq^x_j = \sgm^x_j\sgm^z_{2j+1},\quad
j=1,\ldots,2^{\Lvt-1}-1. 
\end{equation}
It should be noted, that $\btq^x_j$ and $\btq^y_j$ in \Eq{qgenxy} correspond to 
links marked by exchanged labels ($y$ and $x$ respectively, see \Fig{binDFA}).
Remaining $z$-elements \Eq{qgen} can be assigned to `forks' with both links   
\begin{equation}\label{qgenz}
\btq^z_j = \sgm^z_j\sgm^z_{2j}\sgm^z_{2j+1},\quad 
j=1,\ldots,2^{\Lvt-1}-1. 
\end{equation}
Due to Lie-algebraic approach the linear combinations
of quadratic expressions such as \Eq{qgen} correspond to the Hamiltonians ${\xtop H}$ and
the quantum gates can be represented as exponents
\begin{equation}\label{qUH}
 {\xtop U} = e^{-\mi{\xtop H}\tau} = \exp\Bigl(\tau\sum_{j<k} h_{jk} \btg_j\btg_k\Bigr).
\end{equation}

The Hamiltonians such as \Eq{qgeno} and \Eq{qgenxy} generate one- and two-qubit gates and produce
non-universal set of quantum gates for representation of Spin group corresponding to \Eq{qUH}.
The arbitrary one-qubit gates may be generated by such a way for all terminal qubit nodes due to 
\Eq{qgeno}, but two-qubit gates defined on all links of binary qubit tree are restricted by single-parameter
families with Hamiltonians from \Eq{qgenxy}.  

\section{Annihilation and creation operators}
\label{Sec:Ann}

Let us split $2m$ generators $\e_j$ of some Clifford algebra $\UCl(2m,\CC)$ into
two parts with $m$ elements $\e'_j$, $\e''_j$ to introduce {\em annihilation and creation} 
(`ladder') operators
\begin{equation}\label{ladd}
 \an_j=\frac{\e'_j+\mi \e''_j}{2\mi},\quad \an_j^\dag=\frac{\e'_j-\mi \e''_j}{2\mi},\qquad
 j=1,\ldots,m.
\end{equation}
Due to \Eq{ClAR} the elements satisfy {\em canonical anticommutation relations} (CAR)
\begin{equation}\label{CAR}
\{\an_j,\an_k\} = \{\an_j^\dag,\an_k^\dag\}=0, \quad \{\an_j,\an_k^\dag\} = \delta_{jk}\Id,
\end{equation}
where $j,k = 1,\ldots,m$.

For standard representation of Clifford algebra mentioned earlier \Eq{Clgen}
only first $2m$ generators may be used $\e'_j=\e_{2j-1}$, $\e''_j=\e_{2j}$
and thus
\begin{equation}\label{JWladd}
\an_j = \sgm_1^z\cdots\sgm_{j-1}^z\,\oa_j,\quad
 \an_j^\dag = \sgm_1^z\cdots\sgm_{j-1}^z\,\oa_j^\dag,
\end{equation}
where $j=1,\dots,m$ and $\oa$, $\oa^\dag$ are $2\times 2$ matrices 
\begin{equation}\label{ac2x2}
\begin{array}{lcccl}
 \oa &=& \displaystyle\frac{\sgm^x+\mi\sgm^y}{2} 
 &=& \begin{pmatrix}0&1\\0&0\end{pmatrix}, \\  \\
 \oa^\dag &=& \displaystyle\frac{\sgm^x-\mi\sgm^y}{2}
 &=& \begin{pmatrix}0&0\\1&0\end{pmatrix} 
\end{array}
\end{equation}
with index $j$ is for position in tensor product, {\em i.e.},
$$\oa_j \equiv {\underbrace{\Id\otimes\cdots\otimes\Id}_{j-1}\,}%
\otimes \oa \otimes \underbrace{\Id\otimes\cdots\otimes\Id}_{m-j}.$$
Usual Jordan--Wigner transformation \cite{JW} corresponds to standard representation \Eq{Clgen}.

Let us also introduce analogue notation $\op{n}_k$, $\onz_k$, where
\begin{equation}\label{nn0}
\begin{array}{ll}
\op{n} = \oa^\dag\oa = \displaystyle\frac{\Id-\sgm^z}{2} 
& = \begin{pmatrix}0&0\\0&1\end{pmatrix},\\ \\
\onz = \oa\oa^\dag = \displaystyle\frac{\Id+\sgm^z}{2}
& = \begin{pmatrix}1&0\\0&0\end{pmatrix}. 
\end{array}
\end{equation}

Sometimes in physical applications the ladder operators may be considered as primary 
objects and expressions for generators follow directly from \Eq{ladd}
\begin{equation}\label{elad}
\e'_j = \mi(\an_j+\an_j^\dag), \quad 
\e''_j = \an_j-\an_j^\dag.
\end{equation}
The generators $\e_j$ itself due to such representation also often treated as
creation operator for particle coinciding with own antiparticle, {\em e.g.}, 
{\em Majorana mode} \cite{BK00,Wilc09}.

The ladder operators also can be used to express specific subgroup of Spin group
corresponding to some of quantum gates generated by {\em restricted} set of quadratic 
Hamiltonians \cite{Vla18,TD2}. 
Let us introduce notation 
\begin{equation}\label{aquad}
 \op\Sym_{j,k} = \frac{\an_j^\dag\an_k+\an_k^\dag\an_j}{2},\quad
 \op\Asym_{j,k} = \frac{\an_j^\dag\an_k-\an_k^\dag\an_j}{2\mi}.
\end{equation}
For `vacuum' state
\begin{equation}\label{vac}
 \ket{\vac} \equiv \ket{\underbrace{00\ldots 0}_m},
\end{equation}
$\an_k\ket{\vac} = 0$ and thus, $\op\Sym_{j,k}\ket{\vac} = \op\Asym_{j,k}\ket{\vac} = 0$.
Any Hamiltonians $\op\spH$ expressed as linear combinations of \Eq{aquad}
also has the same property $\op\spH \ket{\vac} = 0$ and 
quantum gate generated by such Hamiltonian for some parameter $\tau$
\begin{equation}\label{gtU}
 \op\spU = \exp(-\mi \op\spH \tau)
\end{equation}
does not change vacuum state $\op\spU \ket{\vac} = \ket{\vac}.$

Let us for certainty suppose consequent indexes $1 \le j < k \le m$ in \Eq{aquad}
with special notation for {\em `occupation number'} operators $\num_k$ and
{\em number of `particles'} (units in computational basis) operator $\Num$
\begin{equation}\label{Num}
 \num_k = \op\Sym_{k,k} = \an_k^\dag\an_k, \quad
 \Num = \sum_{j=k}^m \num_k.
\end{equation}
An important property of the operator \Eq{Num} can be derived directly
from the definition and \Eq{CAR}
\begin{equation}\label{pmNum}
\begin{split}
 \Num\an_j = \an_j\Num - \an_j &= \an_j\,(\Num-\Id),\\
 \Num\an_j^\dag &= \an_j^\dag\,(\Num+\Id).
\end{split} 
\end{equation}
Here again $\Num\ket{\vac} = 0$
and for states such as
\begin{equation}\label{StN}
\begin{split}
\ket{\St^{(N)}_{j_1\ldots j_N}} = \underbrace{\an_{j_N}^\dag \cdots \an_{j_1}^\dag}_N\ket{\vac}&,
\\ 1 \le j_1 < \cdots < j_N \le m&
\end{split}
\end{equation}
from consequent application of \Eq{pmNum} for all $\an_j^\dag$ 
it follows
\begin{equation}\label{NumN}
 \Num\ket{\St^{(N)}_{j_1\ldots j_N}} = N\ket{\St^{(N)}_{j_1\ldots j_N}}. 
\end{equation}
It may be also derived from \Eq{pmNum} or checked directly that quadratic operators \Eq{aquad}
commute with $N$
\begin{equation}\label{comN}
 \op\Sym_{j,k}\Num = \Num\op\Sym_{j,k}, \quad \op\Asym_{j,k}\Num = \Num \op\Asym_{j,k}.
\end{equation}
The Hamiltonians $\op\spH$ with linear combination of terms \Eq{aquad}
also commute with $\Num$ and quantum gates $\op\spU$ generated 
by $\op\spH$ \Eq{gtU} respect subspaces composed from states \Eq{StN}.
Such {\em restricted case} was introduced initially in \citeref{TD2}
and later discussed as a basic example in \citeref{Vla18}.

With standard representation \Eq{JWladd} expression for $\Num$ \Eq{Num}  may be rewritten
\begin{equation}\label{Nz}
 \Num = \Num^z \doteq \sum_{k=1}^m\num_k = \sum_{k=1}^m \frac{\Id-\sgm^z_k}{2} 
= \frac{m}{2}\Id-\frac{1}{2}\sum_{k=1}^m \sgm^z_k
\end{equation}
and eigenvalues $N$ \Eq{NumN} of the operator correspond to number of units in 
computational basis, {\em e.g.}, for $N^z=1$ there are $m$ states
\begin{equation}\label{cr1}
\ket{\St^{(m)}_k} = \an_k^\dag\ket{\vac} = \ket{\un{k}}, 
\end{equation}
where, for standard (Jordan--Wigner) representation
\begin{equation}\label{un1}
	\ket{\un{k}} = \ket{\underbrace{0\dots0}_{k-1} 1 \underbrace{0\dots0}_{m-k}},
	\quad k = 1,\dots,m
\end{equation}
with only unit in position $k$ of the computational basis state, but analogue constructions 
even for binary tree discussed below are more complicated.

\medskip

Let us now introduce similar constructions for {\em binary tree}. 
The indexation \Eq{bgind} is used further with first element $\btg_1$ is dropped
and the \Eq{ladd} is applied to partition 
$\e'_j=\btg_{2j}$, $\e''_j=\btg_{2j+1}$, $j=1,\ldots,m$.
Let us also introduce slightly different notation for binary tree 
ladder operators
\begin{equation}\label{bladd}
\ant_j=\frac{\btg_{2j}+\mi \btg_{2j+1}}{2\mi},\quad 
\ant_j^\dag=\frac{\btg_{2j}-\mi \btg_{2j+1}}{2\mi}
\end{equation}
with $j=1,\ldots,m$.

Only for terminal nodes $j=2^{\Lvt-1},\ldots,2^\Lvt-1$ of binary tree with given $\Lvt$
the operators \Eq{bladd} have more usual form with tensor product of only $2 \times 2$ 
matrices similarly with \Eq{JWladd}. Let us consider simple example with 
 $\Lvt=2$ \Eq{bgind2} and first node $j=1$ is not terminal
\begin{equation}\label{ant1}
\ant_1 = \frac{\sgm_1^x\sgm_2^z+\mi\sgm_1^y\sgm_3^z}{2},\quad
\ant_1^\dag = \frac{\sgm_1^x\sgm_2^z-\mi\sgm_1^y\sgm_3^z}{2}.
\end{equation}
Other operators for $\Lvt=2$ corresponds to terminal nodes with simpler expressions 
\begin{equation}\label{ant23}
\begin{split}
\ant_2 & =\frac{\sgm_1^x\sgm_2^x+\mi\sgm_1^x\sgm_2^y}{2}=\sgm_1^x \oa_2, \\ 
\ant_3 & =\frac{\sgm_1^y\sgm_3^x+\mi\sgm_1^y\sgm_3^y}{2}=\sgm_1^y \oa_3.
\end{split}
\end{equation}
The expressions for operators $\ant_j^\dag$ are complex conjugations of matrices 
and often omitted further.
Let us rewrite \Eq{ant1} using projectors \Eq{nn0}
\begin{align}\label{Ant1}
\ant_1 & = \sgm_1^x(\onz_2-\op{n}_2)(\onz_3+\op{n}_3) + 
 \mi\sgm_1^y(\onz_2+\op{n}_2)(\onz_3-\op{n}_3) \notag\\
 &= \oa_1\onz_2\onz_3 + \oa_1^\dag\onz_2\op{n}_3
 - \oa_1^\dag\op{n}_2\onz_3 - \oa_1\op{n}_2\op{n}_3 . 
\end{align}
The expression correspond to {\em `conditional' annihilation and creation operators} on first
qubit controlled by pair of other qubits.
More general case discussed below for $\Lvt\ge 2$ and $j\ge 1$ is quite similar
with appropriate indexes substituted instead of $1,2,3$ in \Eq{Ant1}.

\begin{figure}[htb]
	\begin{center}	
		\includegraphics[scale=0.75]{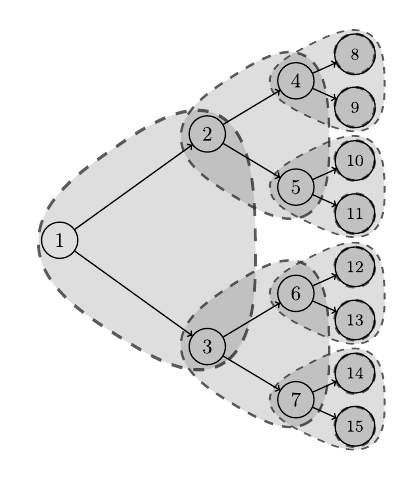}		
	\end{center}
	\caption{Nodes groups for $\ant_j$, $\ant_j^\dag$ in binary tree}
	\label{Fig:bin-sel}
\end{figure}

Let us rewrite \Eq{bladd} with two ranges for internal and terminal nodes
using {\em stub operator} $\stub_j$ \Eq{binstub} together with \Eq{bgind} and \Eq{bgind2} 
\begin{subequations}\label{rladd}
	\begin{align}
	\ant_j & = \stub_j\,\frac{\sgm_j^x\sgm_{2j}^z+\mi\sgm_j^y\sgm_{2j+1}^z}{2}
	= \stub_j \anx{j}{2j}, \notag\\ &
	j=1,\ldots,2^{\Lvt-1}-1,\label{rlad1}\\
	\ant_j & = \stub_j\,\frac{\sgm_j^x+\mi\sgm_j^y}{2} = \stub_j \oa_j,\notag\\ &
	j=2^{\Lvt-1},\ldots,2^\Lvt-1,
	\label{rlad2}   
	\end{align}
\end{subequations}
where $\anx{j}{2j}$ is generalization of conditional operator \Eq{Ant1} 
with index $j$ `controlled' by pair $2j,2j+1$ 
\begin{align}\label{ax}
 \anx{j}{2j} &\doteq \frac{\sgm_j^x\sgm_{2j}^z+\mi\sgm_j^y\sgm_{2j+1}^z}{2}\notag\\
  &= \oa_j(\onz_{2j}\onz_{2j+1} - \op{n}_{2j}\op{n}_{2j+1}) \\
  & + \oa_j^\dag(\onz_{2j}\op{n}_{2j+1} - \op{n}_{2j}\onz_{2j+1}). \notag 
\end{align}
An example for $\Lvt=4$ is depicted on \Fig{bin-sel}. 
The constructions of $\ant_j$, $\ant_j^\dag$ include
three different nodes for $j = 1,\dots,7$ and only one for
$j = 8,\dots,15$.

Let us now consider analogues of \Eq{aquad} 
\begin{equation}\label{quadx}
\xt\Sym_{j,k} = \frac{\ant_j^\dag\ant_k+\ant_k^\dag\ant_j}{2},\quad
\xt\Asym_{j,k} = \frac{\ant_j^\dag\ant_k-\ant_k^\dag\ant_j}{2\mi}.
\end{equation}
and \Eq{Num} for {\em modified number (of `particles') operator}
\begin{equation}\label{xNum}
\xNum = \sum_{k=1}^m \ant_k^\dag\ant_k = \sum_{k=1}^m\xnum_k,
\end{equation}
where $\xnum_k = \ant_k^\dag\ant_k$ are {\em modified `occupation number' operators}.

The `vacuum state' \Eq{vac} for binary tree also satisfies
$\ant_j\ket{\vac} = 0$ for any $j$. It is clear for terminal nodes $j \ge 2^{\Lvt-1}$,
because tensor product for $\ant_j$ includes $\oa_j$ \Eq{rlad2}. 
For alternative expression with three nodes \Eq{rlad1} 
the controlled terms $\anx{j}{2j}$ \Eq{ax} for $\ket{\vac}$ also act 
as annihilation operator on qubit $j$, because two `control qubits' $2j$ 
and $2j+1$ are zeros, {\em cf}\/ \Eq{Ant1}.

Thus, operators \Eq{Num} also satisfy condition $\xt\Sym_{j,k}\ket{\vac} = \xt\Asym_{j,k}\ket{\vac} = 0$
and the same is true for Hamiltonians represented as linear combination of the operators,
$\xt{\spH} \ket{\vac} = 0$. Quantum gates and circuits generated with such Hamiltonians
\begin{equation}\label{xgtU}
\xt\spU = \exp(-\mi \xt{\spH} \tau)
\end{equation}  
do not change `vacuum state' $\xt\spU \ket{\vac} = \ket{\vac}$ similarly
with $\op\spU$ in \Eq{gtU}, but must commute with modified operator $\xNum$
instead of $\Num$.
 
Let us consider analogues of states \Eq{StN}
\begin{equation}\label{xStN}
\begin{split}
&\ket{\xSt^{({\xt N})}_{j_1\ldots j_{\xt N}}} = \ant_{j_{\xt N}}^\dag \cdots \ant_{j_1}^\dag\ket{\vac},\\
\xNum&\ket{\xSt^{({\xt N})}_{j_1\ldots j_{\xt N}}} = \xt N\ket{\xSt^{({\xt N})}_{j_1\ldots j_{\xt N}}}. 
\end{split}
\end{equation}
Quantum gates defined by \Eq{xgtU} due to property $\xNum\xt\spU=\xt\spU\xNum$ do not 
change $\xt{N}$, but number of units in elements of computational basis may be not fixed.

Let us consider example of \Eq{xStN} with single creation operator
\begin{equation}\label{xSt1}
\ket{\xSt_k^{(1)}} = \ant_k^\dag\ket{\vac} \doteq \ket{\xst{k}},
\quad  \xNum\ket{\xst{k}} = \ket{\xst{k}},
\quad  1 \le k \le m.
\end{equation}
The operators $\ant_k^\dag$ are obtained from $\ant_k$ \Eq{rladd} by Hermitian conjugation
and $\ket{\xst{k}}$ is up to phase $\pmi$ 
an element of computational basis with units only in positions corresponding to `path' 
from root to node $k$. The number of units is equal to level $\ell$ of the node in the tree
\begin{equation}\label{levj}
 \Num^z \ket{\xst{k}} = \ell_k \ket{\xst{k}}, \quad \ell_k = \lfloor \log_2 k \rfloor + 1 .
\end{equation}

The eigenvalues of $\xNum$ operators \Eq{xNum} can be expressed directly for computational basis
using analogue of sums \Eq{Num} or \Eq{Nz} with operators $\xnum_j$ written for different 
ranges using \Eq{rladd}
\begin{subequations}\label{xnum}
	\begin{align}
	\xnum_j = \ant_j^\dag\ant_j & = \frac{\Id-\sgm_j^z\sgm_{2j}^z\sgm_{2j+1}^z}{2}, \notag\\
	 &j =1,\ldots,2^{\Lvt-1}-1,\label{xnum1}\\
	\xnum_j =\ant_j^\dag\ant_j & = \frac{\Id -\sgm_j^z}{2},\notag\\
	 &j =2^{\Lvt-1},\ldots,2^\Lvt-1.
	\label{xnum2}   
	\end{align}
\end{subequations}
With quadratic expressions $\btq$ defined earlier \Eq{qgen} it may be rewritten
using \Eq{qgeno} and \Eq{qgenz}
\begin{equation}\label{xqnum}
	\xnum_j = \frac{\Id-\btq_j^z}{2},
	\quad j=1,\ldots,2^{\Lvt}-1.
\end{equation} 
Tensor product of $\sgm^z$ is diagonal matrices and eigenvalues $\eta_j$ of $\btq^z_j$ \Eq{qgenz}
for eigenvectors from computational basis can be expressed as
\begin{multline}\label{ein3}
\btq^z_j\ket{n_1,\ldots,n_m} = \eta_j\ket{n_1,\ldots,n_m}, \\
\eta_j = (-1)^{n_j+n_{2j}+n_{2j+1}} \quad
(j < 2^{\Lvt-1})
\end{multline}
and due to simple identity $$\frac{1-(-1)^k}{2} = k \bmod 2$$ 
eigenvalues of $\xnum_j$ using \Eq{xnum} and \Eq{ein3} can be expressed as
\begin{equation}\label{xtn}
 \xt{n}_j = \begin{cases}
               n_j \exor n_{2j} \exor n_{2j+1}, & j=1,\ldots,2^{\Lvt-1}-1,\\
               n_j,        & j=2^{\Lvt-1},\ldots,2^\Lvt-1,
            \end{cases}
\end{equation}
where $\exor$ denotes {\sf XOR} (exclusive {\sf OR}) operation for binary values
\begin{equation}\label{xn3x}
	\xt n_j = n_j \exor n_{2j} \exor n_{2j+1} = (n_j + n_{2j} + n_{2j+1}) \bmod 2.
\end{equation} 
The eigenvalue of $\xNum$ is
\begin{equation}\label{xtN}
 \xt N = \sum_{j=1}^m \xt{n}_j.
\end{equation}

Let us consider an example with single creation operator for node $k$ \Eq{xSt1}.
The positions of units produce some path from root to $k$.  
Any triple of nodes in \Eq{xtn} for $j \neq k$ contains zero or two units and 
$\xt{n}_j$ is only nonzero element in sum \Eq{xtN}, $\xt{n}_j=\delta_{jk}$, 
thus, $\xt N = 1$.

Let us consider $m$ elements with a single unit in computational basis.
Method used above illustrates that $\xt N = 1$ only for $j=1$, but $\xt N = 2$ for 
$j > 1$ due to second unit in sum \Eq{xtN}, 
because triple for $k = j \mdiv 2$ in \Eq{xtn} also contains node $j$.
It may be also checked directly, that for given indexing \Eq{bgind}
\begin{equation}\label{xSt2x}
\begin{split}
 \ket{\xSt^{(2)}_{j'\!,j}} =  \ant_j^\dag\ant_{j'}^\dag\ket{\vac}, 
 \quad j &= 2,\dots,m = 2^\Lvt-1, \\ j' &= j \mdiv 2
\end{split} 
\end{equation}
is element of computational basis (up to $\pmi$) 
with single unit in position $j$, see \Eq{un1}
\begin{equation}\label{xSt1pos}
\ket{\xSt^{(2)}_{j \div 2,j}} = \ant_j^\dag\ant_{j \div 2}^\dag\ket{\vac} = \pmi\ket{\un{j}},  
\quad j = 2,\dots,m,
\end{equation}
there notation $j \div 2 = j \mdiv 2$ is used for brevity and 
both elements in each pair $j \in \{2j',2j' + 1\}$ are taken into account
for $j>1$. Thus
\begin{equation}\label{unxNum}
 \xNum\ket{\un{1}} = \ket{\un{1}}, \quad
 \xNum\ket{\un{j}} = 2 \ket{\un{j}}, \quad
 \quad j > 1.
\end{equation}
However, elements of computational basis with units in both positions 
$2j'$ and $2j' + 1$ also may be expressed in similar way
\begin{equation}\label{xSt2pos}
\begin{split}
\ket{\xSt^{(2)}_{2j',2j'+1}} &= \ant_{2j'+1}^\dag\ant_{2j'}^\dag\ket{\vac} 
= \ket{\un{2j',2j'+1}}, \\ j' &= 1,\dots,2^{\Lvt-1}-1,
\end{split}
\end{equation}
where notation from \citeref{Vla18} is used
\begin{equation}\label{un2}
\ket{\un{k,k+1}} = \ket{\underbrace{0\dots0}_{k-1} 1 1 \underbrace{0\dots0}_{m-k-2}}.
\end{equation}
Thus, such a states also belong to subspace corresponding to eigenvalue $2$ of operator $\xNum$,
{\em cf} \Eq{unxNum}
\begin{equation}\label{unxNum2}
\xNum\ket{\un{2j,2j+1}} = 2 \ket{\un{2j,2j+1}}, 
\quad 1 \le j \le 2^{\Lvt-1}-1.
\end{equation}

Let us recollect that quantum circuits with gates generated by Hamiltonians \Eq{xgtU} can be used
for transformation between different states from subspaces with {\em the same} eigenvalue of $\xNum$.

\section{Efficient simulation}
\label{Sec:Eff}

Let us start with analogues of efficient classical simulation
considered in \citeref{JM8,JKMW9} with calculation of expectation values
of generators $\btg_j$ for binary trees using exponential representation
of gates ${\xtop U}$ with `quadratic' Hamiltonians ${\xtop H}$ \Eq{qUH}.

Unitary operators $\pm{\xtop U}_{\mt R} \in \mathrm{SU}(2^m)$ (elements of Spin group) 
are corresponding to orthogonal matrix $\mt R$ with property
\begin{equation}\label{xspin}
 {\xtop U}_{\mt R} \btg_j {\xtop U}_{\mt R}^\dag =  \sum_k \mt R_{kj} \btg_k,
\end{equation}
where summation is applied to actually used set of indexes. For binary tree  
natural choice may include either $k = 1,\ldots,2m+1$ for $\UCl(2m+1)$, Spin$(2m+1)$ 
and $\mt R \in \mathrm{SO}(2m+1)$
or $k=2,\ldots,2m+1$ for $\UCl(2m)$, Spin$(2m)$ and 
$\mt R \in \mathrm{SO}(2m) \subset \mathrm{SO}(2m+1)$,
{\em cf\/} \Eq{bgind2} for $m=3$.

Here consideration of all generators with $\mt R \in \mathrm{SO}(2m{+}1)$ 
may be useful, because $\btg_1$ appears in quadratic Hamiltonian
in terms for links such as $\btq_1^x$,  $\btq_1^y$ in \Eq{qgenxy}.
However, $\btg_1$ is dropped in constructions with creation and annihilation operators \Eq{bladd}. 

Evolution of state due to such unitary operators is $\ket{\phi'} = {\xtop U}_{\mt R}\ket{\phi}$
and expectation value of $\btg_j$ is  
\begin{equation}\label{xpbtr}
\bra{\phi'}\btg_j\ket{\phi'} = \bra{\phi}{\xtop U}_{\mt R}^\dag\btg_j{\xtop U}_{\mt R}\ket{\phi} 
= \sum_k \mt R_{jk} \bra{\phi}\btg_k\ket{\phi},
\end{equation}
where order of indexes is changed in comparison with \Eq{xspin} due to inversion 
${\xtop U}_{\mt R}^\dag = {\xtop U}_{\mt R}^{-1}$.
\Eq{xpbtr} is the formal algebraic analogue of an equation for matchgates \cite{JM8}
with $\mt R \in \mathrm{SO}(2m)$, but for the different operators 
${\xtop U}_{\mt R}$, $\btg_j$ are constructed using binary trees instead of linear chain. 
The quadratic terms were more suitable in \citeref{JM8,JKMW9} and analogues of such expressions
also can be introduced
\begin{align}\label{xpqbtr}
\bra{\phi'}\mi\btg_{j_1}\btg_{j_2}\ket{\phi'} &= \bra{\phi}\mi{\xtop U}_{\mt R}^\dag\btg_{j_1}\btg_{j_2}{\xtop U}_{\mt R}\ket{\phi}\notag \\
& = \bra{\phi}\mi({\xtop U}_{\mt R}^\dag\btg_{j_1}\xt U_{\mt R})({\xtop U}_{\mt R}^\dag\btg_{j_2}{\xtop U}_{\mt R})\ket{\phi}\\\notag 
&= \sum_{k_1 \neq k_2} \mt R_{k_1 j_1}\mt R_{k_2 j_2} \bra{\phi}\mi\btg_{k_1}\btg_{k_2}\ket{\phi},
\end{align}
where condition $k_1 \neq k_2$ can be used because terms with equal indexes
are disappear due to orthogonality of matrix ${\mt R}$.

For {\em terminal} indexes $j=2^{\Lvt-1},\ldots,2^\Lvt-1$ quadratic terms 
$\mi\btg_{2j}\btg_{2j+1} = \btq^z_j$ \Eq{qgen}
are equal with single Pauli matrix $\sgm^z_j$  \Eq{qgeno}  and expectation value
is analogue with \citeref{JM8,JKMW9}.
However, for {\em internal} indexes  $j=1,\ldots,2^{\Lvt-1}-1$, $\btq^z_j$ are product of 
three Pauli matrices \Eq{qgenz}. 
It may be written
\begin{equation}\label{xavz}
	 \bra{\phi}\mi\btg_{2j}\btg_{2j+1}\ket{\phi} 
	  = \begin{cases}
		\bra{\phi}\sgm^z_j \sgm^z_{2j} \sgm^z_{2j+1}\ket{\phi}, \\ \qquad j=1,\ldots,2^{\Lvt-1}-1\\ \\
		\bra{\phi}\sgm^z_j\ket{\phi},\\   \qquad j=2^{\Lvt-1},\ldots,2^\Lvt-1
	\end{cases}.
\end{equation}
Using definition of $\xnum_j$ \Eq{xqnum} it may be rewritten in agreement with analogue 
equation for $\xt n_j$ \Eq{xtn}
\begin{equation}\label{xavn}
   \bra{\phi}\xnum_j\ket{\phi} = \Av{\xt n_j}
	= \begin{cases}
		\Av{n_j \exor n_{2j} \exor n_{2j+1}}, \\ \ \qquad j=1,\ldots,2^{\Lvt-1}-1\\ \\
		\Av{n_j},\   j=2^{\Lvt-1},\ldots,2^\Lvt-1
	\end{cases},
\end{equation}
where notation $\Av{\cdots}$ for expectation value is used, 
{\em e.g.}, $\Av{n_j}={p_1}_j$ is probability to measure value $1$ for qubit $j$.

For terminal nodes $j=2^{\Lvt-1},\ldots,2^\Lvt-1$ the result of qubit measurement in computational 
basis \mbox{$n_j=\xt n_j$} can be directly found from \Eq{xpqbtr}. 
For previous level $\ell_j=\Lvt-1$ with indexes $j=2^{\Lvt-2},\ldots,2^{\Lvt-1}-1$ it includes
an expression with three terms
$$
 n_j = \xt n_j \exor n_{2j} \exor n_{2j+1} = \xt n_j \exor \xt n_{2j} \exor \xt n_{2j+1},
$$
for level $\ell_j=\Lvt-2$ expression $n$ via $\xt n$ require seven terms
$$
n_j = \xt n_j \exor \xt n_{2j} \exor \xt n_{4j} \exor \xt n_{4j+1} 
              \exor \xt n_{2j+1} \exor \xt n_{4j+2} \exor \xt n_{4j+3}.
$$
For deeper levels $\ell_j=\Lvt-d$ similar expansions produce $2^{d+1}-1$ terms
\begin{equation}\label{invxn}
n_j = \begin{cases}
 \Bigl(\xt n_j + \sum\limits_{k \in d(j)} \xt n_k\Bigr) \bmod 2, \\ \qquad\qquad j=1,\ldots,2^{\Lvt-1}-1\\ \\
\xt n_j, \quad        j=2^{\Lvt-1},\ldots,2^\Lvt-1
\end{cases},
\end{equation}
where $d(j)$ are {\em all descendants} of node $j$, or, more briefly
\begin{equation}\label{invxn'}
n_j = \sum_{k \in s(j)} \xt n_k \mod 2.
\tag{\ref{invxn}$'$}
\end{equation}
where $s(j) = d(j) \cup \{j\}$ are all nodes of subtree with root $j$, including trivial case
with single term $s(j)=\{j\}$ for terminal qubit nodes.

Thus, an analogue of approach used in \citeref{JM8,JKMW9} can be applied
only either to computation of $\Av{\xt n_j}$ or for measurements of 
separate qubits in terminal nodes. 

For internal nodes with level $\ell < \Lvt$ even for single qubit measurement outcome 
should be used more complicated approach similar with applied to multi-qubit 
outputs in a standard case \cite{Br16}, but with measurement of $2^{\Lvt-l}$ 
quantum `binary variables' $\xt n_j$ expressed as {\sf XOR} operations with
qubit values. 
Thus, despite of some resemblance with matchgate circuits the effective modeling 
with binary trees devotes special consideration. 

Together with possible difficulties
for internal nodes it has specific advantages for terminal qubits.
Linear combinations of quadratic Hamiltonians \Eq{qgeno} may generate
arbitrary rotation and expectation values $\Av{Z_j}$ 
in computational basis \Eq{xavz} can be extended for efficient simulation
of qubit measurement `along any axis.'

A pair of terminal qubits with indexes $2j$, $2j+1$ have 
common parent $j=2^{\Lvt-2},\ldots,2^{\Lvt-1}-1$. Let us show,
that for parent qubit fixed in state $\ket{0}$ any 
transformation from SU$(4)$ group may be implemented
using only quadratic Hamiltonian. 
The construction with auxiliary qubit uses isomorphism between SU$(4)$ and Spin$(6)$ 
and similar with a method discussed in \citeref{Vla15}.

Let us extend a simpler example $\Lvt = 2$, $m = 3$ \Eq{bgind2} 
to write seven generators associated with the `terminal triple' of qubits
with parent node $2^{\Lvt-2} \le j < 2^{\Lvt-1}$ for arbitrary $\Lvt \ge 2$  
\begin{alignat}{3}\label{bgindj}
\btg_j = \stub_j\sgm_j^z,\qquad
\btg_{2j} &= \stub_j\sgm_j^x\sgm_{2j}^z,&\quad
\btg_{2j+1} &= \stub_j\sgm_j^y\sgm_{2j+1}^z,\notag\\
\btg_{4j} &= \stub_j\sgm_j^x\sgm_{2j}^x,&
\btg_{4j+2} &= \stub_j\sgm_j^y\sgm_{2j+1}^x,\\
\btg_{4j+1} &= \stub_j\sgm_j^x\sgm_{2j}^y,&
\btg_{4j+3} &= \stub_j\sgm_j^y\sgm_{2j+1}^y.\notag
\end{alignat}
Products of two generators \Eq{bgindj} produces $21$ different terms,
but only $15$ of them do not change parent qubit with state $\ket{0}$
\begin{equation}\label{s6gen}
 \sgm_{2j}^\mu,\quad\sgm_{2j+1}^\nu,\quad
 \sgm_j^z\sgm_{2j}^\mu\sgm_{2j+1}^\nu,\qquad \mu,\nu = x,y,z.
\end{equation}
The linear combinations of analogues of terms \Eq{s6gen} without 
multiplier $\sgm_j^z$ would produce arbitrary traceless Hamiltonian for two qubits, 
but $\sgm^z$ acts as identity on state $\ket{0}$ and so terms \Eq{s6gen} 
also may generate arbitrary SU$(4)$ transformation of two terminal qubits 
if common parent qubit is $\ket{0}$. \qed

\bigskip

Let us now consider construction of gates $\xt\spU$ \Eq{xgtU} generated by
quadratic combinations \Eq{quadx} of ladder operators $\ant_j$ and $\ant_k^\dag$ \Eq{bladd}
for binary tree. For such a case instead of \Eq{xspin} an auxiliary matrix 
$\mt U \in \mathrm{SU}(m)$ can be introduced for operators
$\pm\xsUU \in \mathrm{SU}(2^m)$ with formal analogue
of well-known relations for ladder operators \cite{Vla18,TD2}
\begin{equation}\label{xospin}
\xsUU \ant_k \xsUU^\dag =  \sum_{j=1}^m \mt U_{kj} \ant_j,\quad
\xsUU \ant_k^\dag \xsUU^\dag 
=  \sum_{j=1}^m \mt U^\dag_{jk} \ant^\dag_j,
\end{equation}
where $\mt{\bar U}_{kj}$ is complex conjugation of coefficients and \mbox{$\mt{U^\dag}=\mt{U^{-1}}$} 
for unitary matrix $\mt{U}$.

A `path-state' $\ket{\xst{k}}$ \Eq{xSt1} satisfies an analogue of equations used in \citeref{Vla18}
for $\ket{\un{k}}$ defined by \Eq{cr1} up to trivial change of variables, {\em i.e.},
\begin{equation}\label{Uxst}
\begin{split}
 \xsUU\ket{\xst{k}} &= \xsUU\ant_k^\dag\ket{\vac} = \xsUU\ant_k^\dag\xsUU^\dag\ket{\vac} \\
 &=  \sum_{l=1}^m \mt U^\dag_{lk} \ant_l^\dag \ket{\vac} = \sum_{l=1}^m \mt U^\dag_{lk} \ket{\xst{l}}.
\end{split} 
\end{equation}
Let us consider linear superposition of path states $\ket{\xst{\chi}}= \sum_{k=1}^m\chi_k\ket{\xst{k}}$   
\begin{equation}\label{Uxsup}
\begin{split}
\xsUU\ket{\xst{\chi}} &= \xsUU\sum_{k=1}^m\chi_k\ket{\xst{k}} =
 \sum_{l,k=1}^m \mt U^\dag_{lk} \chi_k\ket{\xst{l}} \equiv
 \sum_{l=1}^m  \chi_l'\ket{\xst{l}},\\
 \chi'_l &\doteq \sum_{k=1}^m \mt U^\dag_{lk} \chi_k .
\end{split} 
\end{equation}

The \Eq{Uxsup} for `single-path' states ($\xt N=1$) is similar with evolution of 
`single-particle' case ($N^z=1$) for {\em qubit chain} \cite{Vla18}, but 
for all nodes except of root in {\em binary qubit tree} $\ket{\un{k}}$ 
belongs to $\xt N=2$ subspace due to \Eq{unxNum}.
However, the same subspace also includes pairs $\ket{\un{2j,2j+1}}$ \Eq{unxNum}
and an analogy with `two-particle' case is also relevant.

For Hamiltonians respecting $\Num$ or $\xNum$ the consideration of `number-preserving' 
subspaces is natural for models of state transfer in quantum chains \cite{Vla18,LAPPP15} or trees. 
The two-qubit state can be decomposed into three parts:
\begin{equation}\label{q2part}
\ket{\psi} = \underbrace{\overbrace{c_{00}\ket{00}}^{N=0}}_{\xt N = 0} 
+ \underbrace{\overbrace{c_{01}\ket{01} + c_{10}\ket{10}}^{N=1}
	+ \overbrace{c_{11}\ket{11}}^{N=2}}_{\xt N = 2}\ ,
\end{equation} 
but terms with $N=1$ and $N=2$ in \Eq{q2part} in binary tree for pairs of
nodes $2j,2j+1$  ($0 < j < 2^{\Lvt-1}$) are belong to the same 
subspace $\xt N=2$, and, furthermore, \mbox{$N=\xt N = 0$} is not affected 
by $\xt\spU$ \Eq{xgtU} for state transfer.

For two consequent indexes $2j, 2j+1$ three terms with $N \neq 0$ ($\xt N=2$) in \Eq{q2part}
are generated by applications to $\ket{\vac}$ different pairs of operators between the same triple 
$\ant_j^\dag$, $\ant_{2j}^\dag$ and $\ant_{2j+1}^\dag$ due to \Eq{xSt1pos} and \Eq{xSt2pos}.
Thus, result of perfect transfer of such two-qubit pair into new position $2k, 2k+1$ indexes 
by operator $\xsUU$ should correspond to unitary matrix $\mt U$ with simple constrains on 
three elements
\begin{equation}\label{Uperf}
 |\mt U_{jk}| = |\mt U_{2j,2k}| = |\mt U_{2j+1,2k+1}| = 1. 
\end{equation}
For consideration of perfect transfer with single qubit one condition in \Eq{Uperf} 
may be superfluous.

The example illustrates possibility of exponential decrease of model 
dimension from $2^m$ to $m$, but the construction of $\mt U$ 
with a sequence of steps or appropriate Hamiltonians devotes separate consideration elsewhere. 
Even reduced problem is more difficult than analogue example for qubit chain because of less 
trivial structure of graph itself and more complicated properties of modified operators such as
$\ant$ and $\ant^\dag$.

\section{General trees}
\label{Sec:gen}

\subsection{Alternative encoding of binary trees}
\label{Sec:altbin}

In the binary trees discussed earlier all nodes attached to $z$-links were deleted. 
Let us consider as an alternative the binary {\it x-z} trees with $y$-links collapsed instead.
The {\em stub operator} $\stub_j$ \Eq{stub} for such a tree contains 
$\sgm^x$, $\sgm^z$ and generators may contain no more than one $\sgm^y$.

\begin{figure}[htb]
	\begin{center}	
		\includegraphics[scale=0.75]{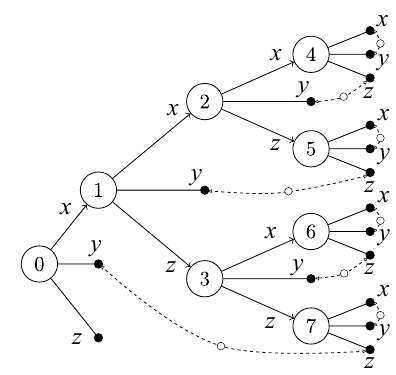}		
	\end{center}
	\caption{Pair of nodes in binary {\it x-z} $\oCFtr{\Lvt}$-tree for $\Lvt=3$}
	\label{Fig:altbin}
\end{figure}

Some constructions discussed below become more natural, if
new root with index zero is attached by $x$-link. 
Similar method was briefly mentioned in \Sec{Tern} and
for $\CFtr{\Lvt}$-tree it produces
`$\oCFtr{\Lvt}$-tree' of height $\Lvt$ with $2^\Lvt$ nodes. 
In such a case appropriate pairs of generators can be chosen to provide 
necessary coupling of $\sgm_x$ and $\sgm_y$ for qubits with 
the same index for specific construction of ladder operators \Eq{ladd}
discussed below, see \Fig{altbin}. 

Let us consider an example with eight qubits.
Similarly with binary trees discussed earlier, only $z$-term for root $\sgm_0^z$ 
is excluded from such a coupling and internal nodes require more complicated 
expressions for ladder operators 
\begin{subequations}\label{altan}
\begin{equation}\label{altint}
\begin{split}
\xz{\am}_0=\frac{\sgm_0^x\sgm_1^z\sgm_3^z\sgm_7^z+\mi\sgm_0^y}2,&\quad
\xz{\am}_1=\sgm_0^x\frac{\sgm_1^x\sgm_2^z\sgm_5^z+\mi\sgm_1^y}2,\\
\xz{\am}_2=\sgm_0^x\sgm_1^x\frac{\sgm_2^x\sgm_4^z+\mi\sgm_2^y}2,&\quad
\xz{\am}_3=\sgm_0^x\sgm_1^z\frac{\sgm_3^x\sgm_6^z+\mi\sgm_3^y}2,
\end{split} 
\end{equation}
in comparison with terminal qubit nodes, {\em cf} \Eq{rlad2}
\begin{equation}\label{altext}
\begin{split}
 \xz{\am}_4=\sgm_0^x\sgm_1^x\sgm_2^x\frac{\sgm_4^x+\mi\sgm_4^y}2,&\quad
 \xz{\am}_5=\sgm_0^x\sgm_1^x\sgm_2^z\frac{\sgm_5^x+\mi\sgm_5^y}2,\\
 \xz{\am}_6=\sgm_0^x\sgm_1^z\sgm_3^x\frac{\sgm_6^x+\mi\sgm_6^y}2,&\quad
 \xz{\am}_7=\sgm_0^x\sgm_1^x\sgm_3^z\frac{\sgm_7^x+\mi\sgm_7^y}2.
\end{split} 
\end{equation}
\end{subequations}

In such construction for each terminal node $j$ there are two generators with
terms $\sgm^x_j$ and $\sgm^y_j$ coupled by natural way \Eq{altext}, but
generator with $\sgm^z_j$ is coupled with some internal node $j'$
linked with $j$ by path $xz\cdots z$ in agreement with \Eq{altint}, see \Fig{altbin}.

Let us consider structure of expressions for internal nodes such as \Eq{altint}. 
For some set of nodes (`chain') $c = \{c_1,\ldots,c_l\}$
the short notation is used further
\begin{equation}\label{prdz}
\prdz c = \sgm^z_{c_1}\cdots\sgm^z_{c_l}.
\end{equation}
Let us also introduce operators
\begin{equation}\label{nopl}
\begin{split}
\op{n}_{\Xr c} &= \frac{\Id - \sgm^z_{c_1}\cdots\sgm^z_{c_l}}{2}
= \frac{\Id - \prdz c}{2}, \\
\onz_{\Xr c} &= \Id -\op{n}_{\Xr c} = \frac{\Id + \prdz c}{2}. 	
\end{split}
\end{equation}
Such projectors have eigenvalues expressed as {\sf XOR}
of nodes from set $c$
\begin{equation}\label{enopl}
\begin{split}
\op{n}_{\Xr c}\ket{n_1,\ldots,n_m} &= n_{\Xr c}\ket{n_1,\ldots,n_m},\\
n_{\Xr c} &= n_{c_1} \exor \cdots \exor n_{c_l}.
\end{split}
\end{equation}
Specific term from expressions for internal nodes such as \Eq{altint}
may be rewritten 
\begin{equation}\label{anpz}
\begin{split}
 \anxr{j}{c} &=\frac{\sgm_j^x \prdz c+i\sgm_j^y}2 \\
 &= \frac{\sgm_j^x +\mi\sgm_j^y}2\cdot\frac{\prdz c + \Id}2 +
\frac{\sgm_j^x -\mi\sgm_j^y}2\cdot\frac{\prdz c - \Id}2 \\
 &=\oa_j\frac{\Id + \prdz c}2 -
 \oa_j^\dag\frac{\Id-\prdz c}2 
 =  \oa_j\onz_{\Xr c} - \oa_j^\dag \op{n}_{\Xr c}.
\end{split} 
\end{equation}
Such a term is an analogue of conditional ladder operator \Eq{Ant1},
because $\anxr{j}{c}$ is also controlled by few nodes $c_1,\dots,c_l \in c$.

The analogue of \Eq{rladd} can be written for binary {\it x-z} $\oCFtr{\Lvt}$-tree
with $2^\Lvt$ nodes taking into account new root with index zero, 
see \Fig{altbin}
\begin{subequations}\label{aladd}
	\begin{align}
    \xz{\am}_j & = \stub_j\,\frac{\sgm_j^x \prdz{c(j)}+\mi\sgm_j^y}{2}
	= \stub_j \anxr{j}{c(j)}, \notag \\ 
	 & j=0,\ldots,2^{\Lvt-1}-1,\label{alad1}\\
	 \xz{\am}_j & = \stub_j\,\frac{\sgm_j^x+\mi\sgm_j^y}{2} = \stub_j \oa_j, \notag\\
	 & j=2^{\Lvt-1},\ldots,2^\Lvt-1, 	\label{alad2}   
	\end{align}
\end{subequations}
where $\stub_j$ is {\em stab operator} already introduced earlier, 
{\em cf\/} \Eq{aladd} for $\Lvt=3$ with \Eq{altan}.
The index $c(j)$ in \Eq{aladd} denotes set of nodes $c_1,\dots,c_l$ attached
to given node $j$ via chain of $z$ links.

The generators of Clifford algebra for \Eq{aladd} in agreement with 
\Eq{elad} can be written
\begin{subequations}\label{agen}
\begin{alignat}{3}
    \xz{\e}'_j & = \mi\stub_j \sgm_j^x \prdz{c(j)},&\quad
    \xz{\e}''_j & = \mi\stub_j\sgm_j^y,&\qquad 
    j\notin \term,\label{agen1}\\
    \xz{\e}'_j & = \mi\stub_j \sgm_j^x,&
    \xz{\e}''_j & = \mi\stub_j\sgm_j^y,& 
   j\in \term. 	\label{agen2}   
   \end{alignat}
\end{subequations}
where $\term$ denotes set of terminal nodes, {\em e.g.,}
$j=2^{\Lvt-1},\ldots,2^\Lvt-1$ for trees used in examples
above.

The analogues of \Eq{xnum} for quadratic operators are also straightforward
\begin{subequations}\label{xznum}
	\begin{alignat}{3}
	\xz{\nm}_j = \xz{\am}_j^\dag\xz{\am}_j & = \frac{\Id-\sgm_j^z\prdz{c(j)}}{2},&\quad
	j\notin \term,\label{xznum1}\\
	\xz{\nm}_j =\xz{\am}_j^\dag\xz{\am}_j & = \frac{\Id -\sgm_j^z}{2},& 
	j \in \term.
	\label{xznum2}   
	\end{alignat}
\end{subequations}

\begin{figure}[htb]
	\begin{center}	
		\includegraphics[scale=0.5]{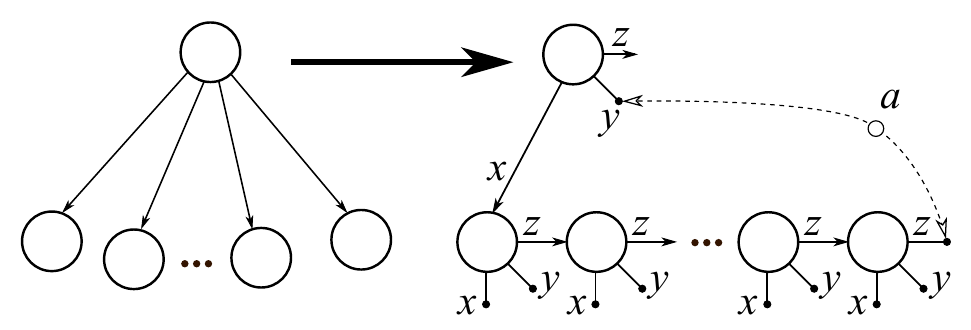}		
	\end{center}
	\caption{Multiple children encoding}
	\label{Fig:nchild}
\end{figure}

The particular example with $2^\Lvt$ nodes is interesting due to direct
relation with Bravyi--Kitaev (BK) transformation discussed below in \Sec{BK}, 
but binary {\it x-z} tree is also can be used to represent 
a general tree ($g$-tree). 
A node $j$ with $l$ children $c_1,\dots,c_l$ of such a $g$-tree 
should be mapped into node $j$ of binary {\it x-z} tree with $x$-link to 
only one child node $c_1$ together with chain of nodes $c_1,\dots,c_l$ 
connected by $z$-links, see \Fig{nchild}. 
For construction of ladder operators the last node $c_l$ 
is coupled with node $j$, {\em cf \/} \Eq{altint}.

Such construction has some properties of formalism used earlier 
due to certain similarity of \Eq{rladd} and \Eq{xnum} for binary {\it x-y} 
trees with \Eq{aladd} and \Eq{xznum} for nodes with arbitrary number of children 
obtained from binary {\it x-z} trees using correspondence depicted on \Fig{nchild}.

An analogue of \Eq{xtn} is
\begin{equation}\label{xztn}
\xz{n}_j = \begin{cases}
n_j \exor n_{c_1} \exor \cdots \exor n_{c_l}, & j \notin \term,\\
n_j,        & j \in \mathcal{T},
\end{cases}
\end{equation}
where  $c_1,\ldots,c_l \in c(j)$
are indexes used in $\prdz{c(j)}$ from \Eq{xznum1}.
It is chain of $z$-linked nodes in node $j$
of initial binary {\it x-z} tree 
and the same indexes correspond to $l$ children of node $j$ in the $g$-tree 
obtained by construction depicted on \Fig{nchild}.

Inverse relation for \Eq{xztn} is similar with \Eq{invxn} used earlier
for binary {\em x-y} trees and may be written
\begin{equation}\label{ixztn}
n_j = \Bigl(\xz n_j + \sum_{k \in D(j)}\xz n_k\Bigr) \bmod 2,
\end{equation}
where $D(j)$ is (possibly empty) set of all descendants of node $j$ for 
$g$-tree obtained from binary {\it x-z} tree. The set of nodes $D(j)$ may differs
from $d(j)$ for corresponding binary {\it x-z} tree, because $z$-link to 
`peers' should not be included in $D(j)$, 
{\em e.g.}, on \Fig{binBK} below $D(3)=\{0,1,2\}$, but $d(3) = \{0,1,2,4,5,6\}$.

\subsection{Bravyi--Kitaev transformation} 
\label{Sec:BK}

Let us compare structure of ladder operators \Eq{aladd} or generators \Eq{agen} with
analogue constructions used in Bravyi--Kitaev transformation based on Fenwick 
trees, see Ref.~\cite{HTW17} and some earlier works \cite{Seel12,Tran15}.
Analogues of operators \Eq{agen} with notation used in \citeref{HTW17} are
\begin{equation}\label{ZYX}
\begin{split}
 \op c_j & = \op Z_{P(j)}\op X_j \op X_{U(j)}, \\
 \op d_j & = \op Z_{C(j)} \op Y_j \op X_{U(j)} = \op Z_{P(j)\setminus F(j)} \op Y_j \op X_{U(j)} 
 \end{split}
\end{equation}
where $\op X$, $\op Y$, $\op Z$ denote either Pauli matrices or theirs products similar with \Eq{prdz},
where  $U(j)$, $C(j)$, $F(j)$ and $P(j) = C(j) \cup F(j)$ are some set of indexes. 
It can be rewritten to provide similarity with notations used here
\begin{equation}\label{szyx}
\xz{\e}'_j  = \mi\prdz{P(j)} \sgm_j^x \prds{x}{U(j)} , \quad
\xz{\e}''_j  = \mi\prdz{C(j)}\sgm_j^y\prds{x}{U(j)},
\end{equation}
where analogue of \Eq{prdz} is used for given set of indexes $S(j)$ and 
Pauli matrix
\begin{equation}\label{prds}
\prds{\mu}{S} = \prod_{s \in S} \sgm^\mu_s.
\tag{\ref{prdz}$'$}
\end{equation}

Thus, operator \Eq{ZYX} from \citeref{HTW17} corresponds to \Eq{agen} 
if $c(j)$ is denoted as $F(j)$ and {\em stub} operator is expressed as
\begin{equation}\label{stubZX}
 \stub_j = \pm \prdz{C(j)}\prds{x}{U(j)}.
\end{equation}

\begin{figure}[htb]
	\begin{center}	
		\includegraphics[scale=0.75]{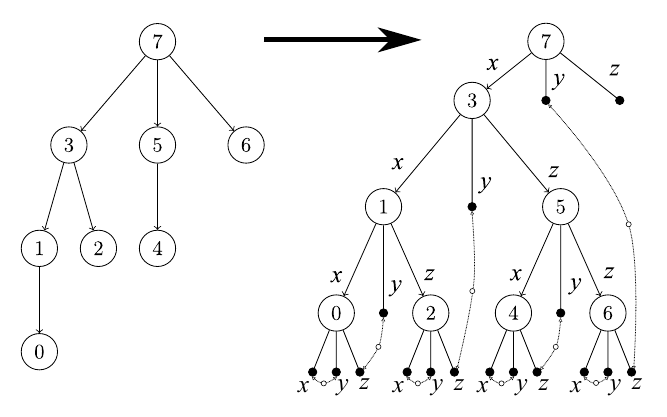}		
	\end{center}
	\caption{Representation of tree used in Bravyi--Kitaev transformation as binary {\it x-z} tree}
	\label{Fig:binBK}
\end{figure}

Let us again consider the example with eight qubits. 
The indexes of nodes in binary {\it x-z} trees should be changed to conform standard
numeration in Bravyi--Kitaev transformation also used in \citeref{HTW17}, {\em cf\/} \Fig{altbin} and \Fig{binBK}
\begin{equation}\label{BKnum}
\begin{array}{|l||c|c|c|c|c|c|c|c|}
\hline
 &0&1&2&3&4&5&6&7\\ \hline
\text{BK} &7&3&1&5&0&2&4&6\\ \hline
\end{array}
\end{equation}

Ladder operators up to numeration \Eq{BKnum} coincide with \Eq{altan} for internal 
\begin{subequations}\label{BKan}
	\begin{equation}\label{BKint}
	\begin{split}
	\xz{\am}_1&=\sgm_7^x\sgm_3^x(\sgm_1^x\sgm_0^z+\mi\sgm_1^y)/2,\\
	\xz{\am}_3&=\sgm_7^x(\sgm_3^x\sgm_5^z\sgm_4^z+\mi\sgm_3^y)/2,\\
	\xz{\am}_5&=\sgm_7^x\sgm_3^z(\sgm_1^x\sgm_2^z+\mi\sgm_1^y)/2,\\
	\xz{\am}_7&=(\sgm_7^x\sgm_3^z\sgm_1^z\sgm_0^z+\mi\sgm_7^y)/2
	\end{split} 
	\end{equation}
and external nodes, respectively	
    \begin{equation}\label{BKext}
	\begin{split}
	\xz{\am}_0&=\sgm_7^x\sgm_3^x\sgm_1^x(\sgm_0^x+\mi\sgm_0^y)/2,\\
	\xz{\am}_2&=\sgm_7^x\sgm_3^x\sgm_1^z(\sgm_2^x+\mi\sgm_2^y)/2,\\
	\xz{\am}_4&=\sgm_7^x\sgm_3^z\sgm_5^x(\sgm_4^x+\mi\sgm_4^y)/2,\\
	\xz{\am}_6&=\sgm_7^x\sgm_3^x\sgm_5^z(\sgm_6^x+\mi\sgm_6^y)/2.
	\end{split} 
	\end{equation}
\end{subequations}

With new indexing \Eq{xztn} may be rewritten for eight qubits depicted on \Fig{binBK}
\begin{equation}\label{BKxz}
\begin{array}{ll}
\xz{n}_0 = n_0,\quad\xz{n}_2 = n_2,&
\xz{n}_4 = n_4,\quad\xz{n}_6 = n_6, \\
\xz{n}_1 = n_1 \exor n_0,&\xz{n}_5 = n_5 \exor n_4, \\
\xz{n}_3 = n_3 \exor n_1 \exor n_2, &
\xz{n}_7 = n_7 \exor n_3 \exor n_5 \exor n_6.\!\!\!
\end{array}
\end{equation}
The inverse relations \Eq{ixztn} are
\begin{equation}\label{BKn}
\begin{array}{ll}
	n_0 = \xz n_0,\qquad n_2 = \xz n_2,&
	n_4 = \xz n_4,\qquad n_6 = \xz n_6, \!\!\!\!\!\\
	n_1 = \xz n_1 \exor \xz n_0,& n_5 = \xz n_5 \exor \xz n_4, \\
	n_3 = \xz n_0 \exor \xz n_1 \exor \xz n_2 \exor \xz n_3, \\
	\multicolumn{2}{l}{
	n_7 = \xz n_0 \exor \xz n_1 \exor \xz n_2 \exor \xz n_3
	\exor \xz n_4 \exor \xz n_5 \exor \xz n_6 \exor \xz n_7. \!\!\!\!\!} 
\end{array}
\end{equation}

Let us recollect, what $n_j$ corresponds to single qubit with index $j$,
but $\xz{n}_j$ is `BK number' related with set of qubits affected by 
`modified BK creation operator' $\xz\am_j^\dag$.

In such a way, the set of equations \Eq{BKn} is in agreement with usual scheme of Bravyi--Kitaev transformation 
\cite{BK00} and it corresponds to an example of Fenwick tree with eight nodes considered in 
\citeref{HTW17} taking into account correspondence between $g$-tree and binary {\it x-z} 
tree discussed in \Sec{altbin}.

\onecolumngrid
\newpage
\section{Conclusion and discussion}
\label{Sec:concl}

\twocolumngrid

Construction of Clifford algebras associated with some kinds of trees is discussed in presented work.
Formally, set of generators can be produced by {\em deterministic finite automaton} obtained as the extension
of ternary tree by addition some formal output nodes. The binary trees can be formally considered as a reduced 
case of ternary tree with at least one child for each node is omitted, see \Fig{ter2bin}.
In appropriate cases the trees can be also used for modeling of quantum state transfer along the edges. 

Spin group can be expressed using exponents with linear combination of terms quadratic by 
generators of Clifford algebra. Such terms correspond to Hamiltonians in quantum mechanics.
The trivial case is a chain associated with standard (Jordan--Wigner) generators of Clifford 
algebra \Eq{Clgen}. In such a case the quadratic expression for Hamiltonian of a node
is $\e_{2k-1}\e_{2k}$ and more general terms $\e_j\e_k$ represent expressions with 
Pauli matrices acting on two or more consequent qubit nodes in the chain.

Both for binary and ternary trees the expressions for generators include sequence of nodes
from root to some terminal node. Thus, quadratic expressions represent single node
or segment with sequence between two nodes. However, number of formal output nodes
of {\em deterministic finite automaton} attached to given qubit is $n_o=3-n_c$, where 
$n_c$ is number of children for given qubit in a tree. Thus, for ternary trees internal 
qubit node may be missing in such sequence and binary trees with $n_o > 0$ are 
more preferable for some purposes.

The construction with trees naturally produces odd number of generators, but any one of them
can be expressed as product of others. Due to such property any generator could be dropped,
yet new set with even number of generators may lack of initial symmetry.
Anyway, even number of generators decomposed on pairs can be used for definition of 
creation and annihilation (ladder) operators \Eq{ladd}. Such construction is appropriate 
for general ternary tree, but it looks more natural for reduced cases such as binary trees
or linear chain.

The generators of Clifford algebra $\e_j$ in some physical applications can be also
treated as creation operators, but particle and antiparticle is equivalent in such
a case, because $\e_j^2=\Id$. The quadratic expressions with generators are convenient
for modeling of state transfer. For a system with $m$ qubits and Hilbert space
with dimension $2^m$ quadratic Hamiltonian produces evolution described by
matrices of rotations in a space with dimension only $2m$ due to main property
of Spin groups \Eq{xspin}.

The \Sec{gen} slightly extends initial topic of this paper about effective
modeling and state transfer to show relations with so-called fermion-to-qubit
mapping for applications in quantum computers. It is shown in \Sec{altbin} that a model with general trees 
often used for such a purposes can be obtained from alternative reduction of ternary 
tree illustrated on \Fig{nchild}. The particular example with Bravyi--Kitaev transformation
is explained in \Sec{BK}.

\end{document}